\documentclass[
  journal=largetwo,
  manuscript=Research article,
  year=2024,
  volume=37,
]{cup-journal}

\usepackage{amsmath}
\usepackage{amssymb}
\usepackage[]{microtype}
\usepackage{booktabs}
\usepackage{xfrac}
\usepackage{hyperref}

\newcommand{\ion}[2]{#1\,{\sc #2}}

\newcommand{\doubletWL}[3]{#1$\lambda\lambda$#2, #3~\AA}
\newcommand{\tripletWL}[4]{#1$\lambda\lambda$#2, #3, #4~\AA}
\newcommand{\quadrupletWL}[5]{#1$\lambda\lambda$#2, #3, #4, #5~\AA}

\newcommand{\HI}{\ion{H}{i}}

\newcommand{\CI}{\ion{C}{i}} \newcommand{\CIp}{[\CI]} 
\newcommand{\CII}{\ion{C}{ii}} \newcommand{\CIIp}{[\CII]}
\newcommand{\CIII}{\ion{C}{iii}} 
\newcommand{\CIV}{\ion{C}{iv}} \newcommand{\CIVp}{[\CIV]}

\newcommand{\NI}{\ion{N}{i}} \newcommand{\NIp}{[\NI]}
\newcommand{\NII}{\ion{N}{ii}} \newcommand{\NIIp}{[\NII]}

\newcommand{\OI}{\ion{O}{i}}\newcommand{\OIp}{[\OI]}
\newcommand{\OIII}{\ion{O}{iii}} \newcommand{\OIIIp}{[\OIII]}

\newcommand{\SII}{\ion{S}{ii}} \newcommand{\SIIp}{[\SII]}
\newcommand{\SIII}{\ion{S}{iii}} 

\newcommand{\NeIV}{\ion{Ne}{iv}} \newcommand{\NeIVp}{[\NeIV]}

\newcommand{\Ha}{$\text{H}_\alpha$}

\newcommand{\CIdoublet}{\doubletWL{\CIp}{4622}{4627}}

\newcommand{\CIVdoublet}{\doubletWL{\CIVp}{5803}{5814}}

\newcommand{\NIdoublet}{\doubletWL{\NIp}{5198}{5200}}
\newcommand{\NIItriplet}{\tripletWL{\NIIp}{6527}{6548}{6583}}

\newcommand{\OItriplet}{\tripletWL{\OIp}{6300}{6364}{6394}}

\newcommand{\OIIItriplet}{\tripletWL{\OIIIp}{4933}{4959}{5007}}

\newcommand{\SIIdoubletMain}{\doubletWL{\SIIp}{6716}{6731}}

\newcommand{\NeIVqaudruplet}{\quadrupletWL{\NeIVp}{4714}{4717}{4724}{4726}}

\newcommand{\be}{\begin{equation}}
\newcommand{\ee}{\end{equation}}

\newcommand{\ba}{\begin{eqnarray}}
\newcommand{\ea}{\end{eqnarray}}

\newcommand{\E}{{\bf E}}

\title{Modelling of the atomic lines' emission of fast moving pulsar nebulae}

\author{I.N. Nikonorov}
\affiliation{Institute of Astronomy, Russian Academy of Sciences, Moscow, 119017 Russia}
\alsoaffiliation{Kazan Federal University, Kazan, 420008 Russia}
\email[I.N. Nikonorov]{inikonorov@inasan.ru}

\author{M.V. Barkov}
\affiliation{Institute of Astronomy, Russian Academy of Sciences, Moscow, 119017 Russia}

\author{M. Lyutikov}
\affiliation{Department of Physics and Astronomy, Purdue University, West Lafayette, IN 47907-2036, USA}


\begin{document}

\begin{abstract}
Bow shocks  generated by pulsars  moving  through weakly ionised interstellar medium  (ISM) produce emission  dominated by  non-equilibrium  atomic transitions.  These  bow shocks are primarily observed as  H$\alpha$  nebulae.
We developed a  package, named {\it Shu}, that calculates non-LTE intensity maps in more than 150 spectral lines, taking into account geometrical properties of the pulsars' motion and  lines of sight.
We argue here that  atomic (\CI, \NI, \OI)  and ionic (\SII, \NII, \OIII, \NeIV) transitions can be used as  complementary and  sensitive probes of ISM. 
We perform self-consistent 2D relativistic hydrodynamic calculations of the bow shock structure and generate non-LTE emissivity maps,  combining global dynamics of  relativistic flows, and detailed calculations of the non-equilibrium ionisation states. We find that though  typically \Ha\ emission is dominant,   spectral  fluxes in \OIIIp,  \SIIp\ and \NIIp\ may become comparable  for relatively slowly moving pulsars. Overall, morphology of non-LTE  emission, especially of the ionic species, is a sensitive probe of the density structures of the ISM.
\end{abstract}

\section{Introduction}
\label{intro}

Pulsars produce  ultrarelativistic winds \citep{1969ApJ...157.1395O,1984ApJ...283..694K}. Their interaction with surrounding media forms a nebula which  shines from radio to hard gamma-rays band \citep{2004vhec.book.....A}.

More than fifty pulsar wind nebulae (PWNe) have been explored in recent decades thanks to the Chandra space telescope \citep{2008AIPC..983..171K}. Among ``Chandra PWN Zoo'' exists a wide class of bow shock nebulae formed by pulsars which left their natal supernova remnants and move with speeds in the range 0~--~1500~km/s, its average value about $\sim$450~km/s \citep{1994Natur.369..127L}. Later, observations in optical spectral lines \citep{2014ApJ...784..154B} reveal direct detection of PWNe bow shocks in inter stellar media (ISM). 

Observations show a formation of nebulae, which have an elongated head-tail structure. Geometrical factors, orientation of the pulsar spin axis relative to its direction of motion, as well as orientation of the line of sight can greatly affect the shape of the head part of the bow shock \citep{2007MNRAS.374..793V,2019MNRAS.484.4760B}. On the other hand, the tail's shape is not affected significantly by the internal properties of the wind. 
 
The shape of the tail can be affected by the following factors:
\begin{enumerate}
        \item \label{hyp:mass_loading} Mass loading process. Dynamics of gas flows are influenced by neutral atoms passing over bow shock and ionising inside nebula, increasing gas particles number density and decreasing temperature \citep{2015MNRAS.454.3886M, 2018MNRAS.481.3394O}.
        
        \item \label{hyp:extern_density_var} Variation of the external density. In consequence, the shock wave spreads with different velocity in different directions and preferably follows a negative gradient of unshocked gas density \citep{2017MNRAS.464.3297Y,2019MNRAS.484.1475T,2020MNRAS.497.2605B}.
\end{enumerate} 

For example, effects of the mass loading may lead to widening of the opening angle of the nebula cone along its length. As the result, it takes  ``head and shoulders'' shape on scales of few -- ten stand-off distances. However,  numerous widenings and necks, sometimes  asymmetric, can only be explained due to variations of the external parameters. 2D hydrodynamic modelling was performed and resulting  \Ha\ emission maps were calculated in \cite{2020MNRAS.497.2605B}. The obtained emission maps in general are consistent with \Ha\ observations of Guitar nebula. In the regions with high density, the Mach cone shock propagation is slower and temperature after the shock is not too high ($T\sim 10^5$~K), so the local plasma emissivity is higher. On emission maps, high density regions look like bright zones in necks. 

In the observed nebulae, a scale of such structures is $\sim 0.1$~pc. Filaments with a similar distance between them were observed in Tycho supernovae remnant (SNR) \citep[][the last one is an alternative model]{2011ApJ...728L..28E,2015ApJ...805..102L}, or in RX~J0852.0-4622 \citep[][]{2012MNRAS.424.3145P}. \cite{2020MNRAS.497.2605B} propose that this observed structures is produced by variation of density in warm ISM surrounding fast moving pulsars. It must have another origin than filaments in cold molecular clouds. In case of cold ISM, the scale is coincident with the Jeans' length, but in case of warm ISM with temperature of 10$^4$~K, the Jeans' length is $\approx$~2~kpc. Consequently, a new process of formation of the structures at the scale 0.1~--~0.3~pc is required (possibly, a specific regime of thermal instability). Calculation of bow-shock PWNe emissivity maps in many spectral lines could allow us to directly compare the modelling results and observational data. 

So far, bow-shock PWNe were systematically observed in \Ha\ line only. The  largest  survey of 9 objects was carried out by \cite{2014ApJ...784..154B}. Spectrum of the head part of PSR J2225+6535 nebula (``Guitar'') only showed hydrogen Balmer series lines \citep{1993Natur.362..133C,2005AstL...31..245L,2013ffep.confE..67D}.  However, for the formation of forbidden lines, the neck structures -- so-called rings -- are expected to be more suitable. These structures are present in many observed PWNe and in favourable conditions could be places of interaction of several shock waves. Such interaction raises density and in consequence the intensity in corresponding lines by a few orders of magnitude. So, the rings could be much brighter in forbidden lines compared to the rest of a nebula. Lines of heavy elements could be even brighter than \Ha.

\cite{2020MNRAS.497.2605B} showed that bow shocks of PWNe highlight inhomogeneities of ISM. With successful detection of bow-shock PWNe in various spectral lines, it will be possible to consider a reconstruction of distribution of the density and the chemical composition of the ISM material around the pulsar. It may also shed light on the formation mechanism of ISM inhomogeneities on ultra-low scales. 

The aim of the present work is to calculate the synthetic emissivity maps of fast-moving PWNe in \Ha\ and various forbidden spectral lines. We developed a package for the calculation of the intensity maps in spectral lines based on hydrodynamic models accounting for gas ionisation state. 

In Section~\ref{sec:methods_models} we introduce the methods used for calculating models and synthetic intensity maps. We also give brief description of the relativistic flow morphology. Section~\ref{sec:maps} is devoted to the visual inspection and the qualitative description of intensity maps. In Section~\ref{sec:Ha_comparison} we compare the overall luminosity, morphology and brightness profiles of the synthetic nebulae with the real ones in \Ha\ line. Also, we test scaling laws of nebulae luminosity and put constraints on some dependencies. In Section~\ref{sec:predictions} we give predictions on most suitable lines and physical conditions to observationally explore bow-shock PWNe and highlight the expected features of the morphology with quantitative estimates given. Section~\ref{sec:conclusion} is dedicated to the general conclusions.

\section{Methods and models}\label{sec:methods_models}
\subsection{Numerical Simulation Setup}

First, 2D relativistic hydrodynamic modelling of pulsar-ISM interaction was performed using the {\it PLUTO} code\footnote{Link http://plutocode.ph.unito.it/index.html} \citep{2007ApJS..170..228M}. In order to simultaneously calculate the hydrodynamic model and calculate the ionisation balance of the plasma in the ISM, we used the {\it MINEq} module \citep{2008A&A...488..429T}. {\it PLUTO} is a modular Godunov-type code entirely written in C and intended mainly for astrophysical applications and high Mach number flows in multiple spatial dimensions. In our simulations we used 3rd order PPM interpolation in space, and 2nd order Runge-Kutta approximation in time with HLLC Riemann solver \citep{2005MNRAS.364..126M}. 

\begin{figure}
        \includegraphics[width=\textwidth,angle=-0, trim=0 1cm 0 2cm]{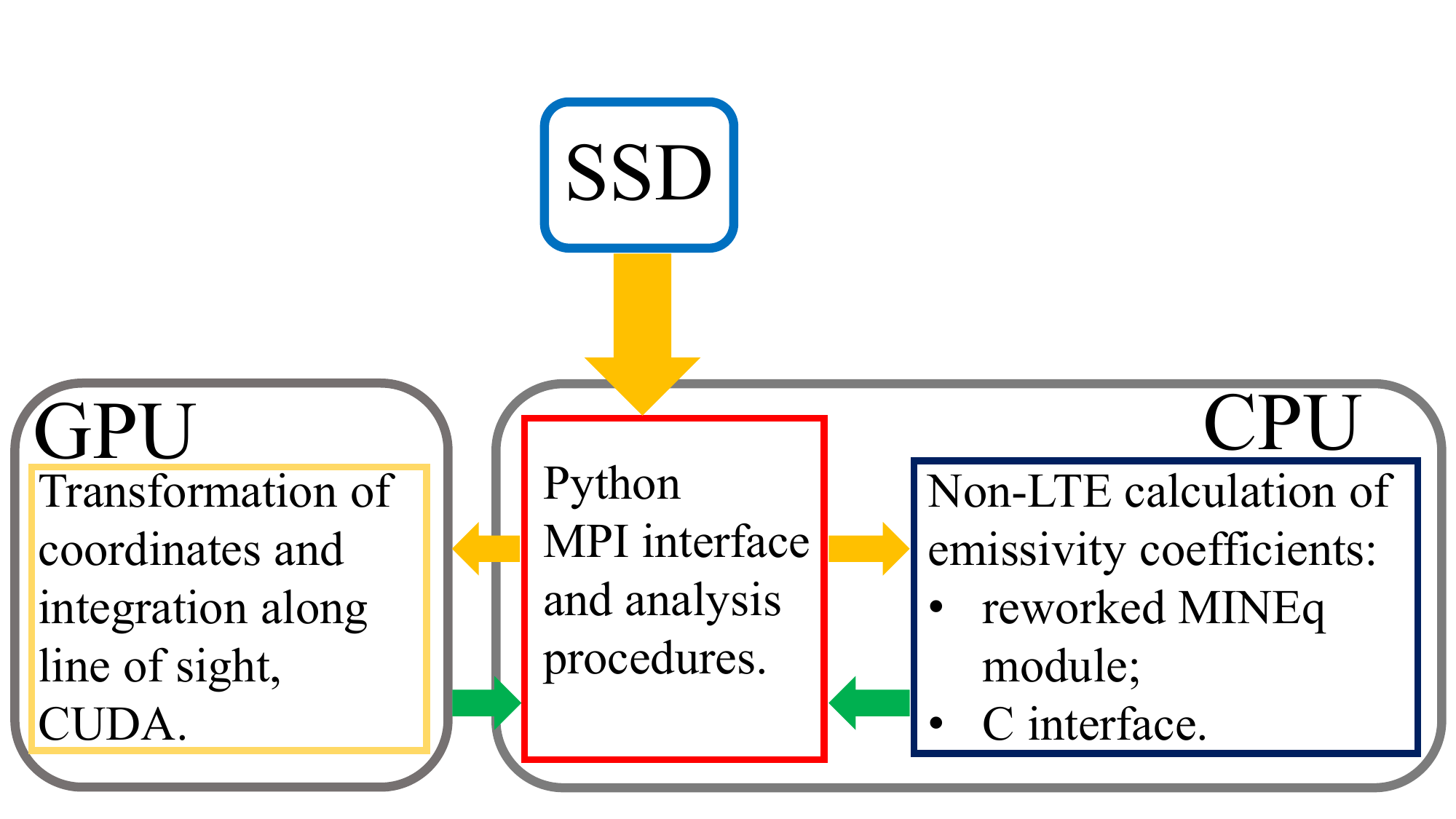}
        \caption{Block-scheme of {\it Shu} package. Yellow arrows show inputs of processes, green ones show outputs.}
        \label{fig:block-scheme}
\end{figure}

Mixing of ions, ionisation, and recombination processes follow the following equation:
\begin{equation}\label{eq:drho1}
        \partial_t (D X_{k,i}) + \nabla \cdot (D X_{k,i} \mathbf{v}) = D S_{k,i}.
\end{equation}
Here, $D$ is a density in the lab frame, $k$ index corresponds to chemical element, $i$ corresponds to the ionisation stage, respectively. $X_{k,i} \equiv n_{k,i} / n_k$ is the fraction of ions, $n_{k,i}$ is the number density of the $i$-th ion of an element $k$, and $n_{k}$ is the density of the element number. $S_{k,i}$ is the source term, which accounts for ionisation, recombination, and radiative energy losses. The conservative part of Equation~\ref{eq:drho1} depicts a transfer of ions and is integrated with hydrodynamic equations.

Only energy density and ionisation states are evolving during cooling, so the action of the $S_{k,i}$ term is described as a system of ODEs:
\begin{equation}\label{eq:ionisation_state_system}
        \frac{d}{dt}
        \begin{pmatrix}
                E \\
                X_{k,i}
        \end{pmatrix}=
        \begin{pmatrix}
                S_E \\
                S_{k,i}
        \end{pmatrix},
\end{equation}
where $S_E$ is a cooling source term in the energy equation.

Solving the ionisation state equations was carried out in the optically-thin plasma limit. The system of Equations~(\ref{eq:ionisation_state_system}) was integrated as a part of {\it MINEq} module apart from hydrodynamics using Runge-Kutta~1-2 method with a switch to Rosenbrock~3-4 if system is stiff and to Cash-Karp~4-5 if estimated error is large.
Joint solution with hydrodynamic equations is achieved with Strang splitting, giving a 3rd order precision in spatial coordinates and 2nd in the temporal coordinate.

The chemical composition of the ISM is shown in Table~\ref{tab:gas_comp}. Atomic data was used from~\citep{2008A&A...488..429T}. For the initial electrons to start ionisation MINEq module has a hard-coded floor of $10^{-4}$ electron per one nucleus in a cell.

\begin{table*}
        \caption{Chemical composition of the gas in models. Here $n_k$ and $n_{at}$ -- number density of $k$-th elements' nuclei and nuclei of all elements respectively.}
        \begin{tabular}{l|ccccccc}
                \hline
                \hline
                Atom ($k$)     & \quad H      & \quad  He     & \quad C                    & \quad N                  & \quad O
                               & \quad Ne     & \qquad S                                                                                                                                                      \\
                \hline
                $n_k / n_{at}$ & \quad $0.93$ & \quad $0.074$ & \quad $3.0 \times 10^{-4}$ & \quad $5 \times 10^{-5}$ & \quad $4.0 \times 10^{-4}$ & \quad $7.0 \times 10^{-5}$ & \qquad $1.5 \times 10^{-5}$ \\
                \hline
        \end{tabular}
         \label{tab:gas_comp}
\end{table*}

Since {\it MINEq} module release dielectronic recombination rates have undergone changes, an overview of most recent sources can be found in \cite{2020ApJ...904..115L}. It would be beneficial to update the data for increase in calculation quality. However, in our setup recombination timescale is much larger than that of ionisation \citep{2020MNRAS.497.2605B}. Overall it weakly influences dynamics and the ionisation state of gas.

In this setup we do not account for photoionisation from X-ray and ultraviolet radiation from neutron star and PWN itself.

\subsection{Grid parameters}

We used a two-dimensional (2D) geometry in cylindrical coordinates. The pulsar is placed in $R=0$ and $z=0$ and the ISM is injected into the computation domain from the left border with speed $V_{NS}$. The unit of length is $a = 10^{16}$~cm as in \cite{2020MNRAS.497.2605B}. The size of the domain is $R \in [0, 35a]$ and $z \in [-3a, 100a]$ ($[-5a, 100a]$ for $V_{NS}=150\text{km/s}$). To have a good resolution in the central region and the long tail zone, we use a non-uniform resolution in the computational domain with the total number of cells $N_{\rm R} = 520$, and $N_{\rm z} = 1560$ ($1690$ for $V_{NS}=150\text{km/s}$). See more details in Table~\ref{tab:grid}. The simulations were performed on CFCA XC50 cluster of National Astronomical Observatory of Japan (NAOJ).

\begin{table*}
        \caption{Parameters of the grid. Here $a=10^{16}$~cm.}
        \begin{tabular}{lccccccc}
                \hline
                \hline
                Coordinates                   & Left, a      & $N_{\rm l}$ & Left-centre, a & $N_{\rm c}$ & Right-centre, a & $N_{\rm r}$ & Right, a     \\
                \hline
                $ R  $                        & \quad —      & \quad —     & \quad $0$      & \quad 130   & \quad $1$       & \quad 390   & \qquad $35$  \\
                $ z  $                        & \quad $-3$   & \quad 130   & \quad $-1$     & \quad 260   & \quad $1$       & \quad 1170  & \qquad $100$ \\
                (for $V_{NS}=150\text{km/s}$) & \quad ($-5$) & \quad (260) & \quad          & \quad       & \quad           & \quad       & \qquad       \\

                \hline
        \end{tabular}
        \label{tab:grid}
\end{table*}

\subsection{Initial and boundary conditions}

Initial and boundary conditions are as follows:
\begin{itemize}
        \item \textit{ISM injection zone}. \\ Velocity of the cold ($p \ll \rho v^2$) gas flow: $(0;~V_{NS})$. Cases:
              \begin{equation}
                      V_{NS}=150,~450,~1500\text{km/s}.
                      \label{eq:vns}
              \end{equation}

              ISM density varies in order to simulate the so-called ``shoulder-neck'' structures in nebulae following the equation:
              \begin{equation}
                      \rho_{ISM} = \frac{\rho_0}{1-a_\rho sin \left[ \frac{z-t V_{NS}}{\lambda}\right]},
                      \label{eq:rhoism}
              \end{equation}
              where $\rho_0=m_p/\text{cm}^3$, $\lambda = 30 \times 10^{16}$cm -- length scale of $\rho_{ISM}$ fluctuations, $a_\rho = 0.5$ -- their amplitude. Equilibrium ionisation state.

              For $V_{NS}=150\text{km/s}$, the ISM density is constant and set to $\rho_0$.

        \item \textit{Internal boundary condition (pulsar wind ejection zone)}. We inject the spherically isotropic cold wind: $p / \rho c^2 = 1 / 100$ (which correspond to Mach number $M=42$), and Lorentz factor is $\Gamma = 4.9$. Stand-off distance (see~\cite{2019MNRAS.484.4760B} for details) is $r_s = 0.57 \times 10^{16}$~cm.

\item \textit{Other boundary conditions} are the following: wind outflow at the tail side of nebula, axisymmetric/fully reflective boundary conditions at central axis, free outflow on outer boundary.

\item For different models (see Table~\ref{tab:models}) we used the ideal equation of state with two values of the adiabatic index 
\begin{equation}
\gamma = \frac{4}{3},~\frac{5}{3}.
\label{eq:gamma}
\end{equation}
For $V_{NS}=150\text{km/s}$ only $\gamma = 4/3$ is considered.

\end{itemize}

All names of the models are listed in Table~\ref{tab:models}:

\begin{table}
        \begin{tabular}{lcccc}
                \hline
                \hline
                Model    & $V_{NS}$,~km/s & $\gamma$  & ISM density    \\
                \hline
                v01g43nv & \quad 150      & \quad 4/3 & \quad Uniform  \\
                v03g43   & \quad 450      & \quad 4/3 & \quad Variable \\
                v03g53   & \quad 450      & \quad 5/3 & \quad Variable \\
                v1g43    & \quad 1500     & \quad 4/3 & \quad Variable \\
                v1g53    & \quad 1500     & \quad 5/3 & \quad Variable \\
                \hline
        \end{tabular}
        \caption{Parameters of the models.}
        \label{tab:models}
\end{table}

\subsection{Method of emissivity calculation}

In order to calculate emissivity maps, a high-performance program package was created. We called it {\it Shu} \citep{Shu2023} after the Egyptian god of the air and supporter of the sky. His ostrich feather was symbolic of lightness and emptiness. {\it Shu} was considered to be a cooling, and thus calming, influence, and a pacifier.

The package uses all resources of a workstation, such as CPU, GPU and fast SSD storage. The structural scheme of the package is shown in Figure~\ref{fig:block-scheme}.

Reading hydrodynamic simulation checkpoints is done in the Python-runtime processes with the PyPLUTO package. Parallelism is implemented on simulation checkpoints using MPI. RAM addresses of the data are being transferred to the C-module based on {\it MINEq}. The major difference between modules is that {\it MINEq} calculates cooling function in the unit of volume, whereas our module calculates an emissivity coefficient:
\begin{align}
        MINEq: S_E            & = - \left( n_{at} n_e \Lambda (T, \mathbf{X}) +L_{FF} +L_{I-R} \right), \label{eq:MINEq_cooling} \\
        \text{Our work}: \eta_{em} & = n_{at} n_e \hat{\Lambda} (T, X_{k,i}) / 4 \pi, 
        \label{eq:PLUTO_emission}
\end{align}
where $\Lambda$ is the part of a cooling function due to cooling in spectral lines, normalised to the concentration of electrons and ions; $\hat{\Lambda}$ -- its part due to lines, which were selected for calculation; $X_{k,i}$ -- the part of $i$-th ion among all of $k$-th element atoms; $\mathbf{X}$ is the gas ionisation state.

After that, arrays of emissivity coefficients are being sent to VRAM by MPI processes. On GPU, a conversion from the non-uniform 2D grid to the uniform 3D grid and the summation along the line of site takes place. The conversion is carried out by the coordinate system rotation and the nearest-neighbour interpolation. Parallelism here is based on the breaking down calculation task to compute individual pixels of the intensity map (about $8\times10^5$ pixels per map, which is much more than number of CUDA cores in one GPU).

Atoms and ions available for calculation are H, He and their ions, five lowest ionisation stages of C, N, O, Ne and S -- 23 species in total. Electron configurations vary from those of hydrogen-like elements ($1s^1$), alkali metals ($2s^1$), helium-like elements ($1s^2$) and alkaline earth metals ($n s^2$) to elements with 1~--~6 p electrons ($n p^1$~--~$n p^6$). The detailed description of the given configurations' spectra can be found in~\cite{1979asrt.book.....S}. The considered configuration belongs to elements in various ionisation stages, generating many emission lines from the ultraviolet to infrared spectral range. At the moment, only configurations with $d$ and $f$ electrons are not available for computation, as well as the fifth ionisation stage, in which an excess of population occurs because of the lack of higher stages.

We studied all optical lines available for the calculation given a restriction of low energy level count in models of ions (usually 3 -- 10 levels including a fine structure), except helium (we plan to upgrade models of H and He atoms and ion and make research on them in a separate work). The goal was to determine which of the factors impact expected observational features of bow-shock PWNe the most. Thus, besides \HI, species with configurations of alkali metals (\CIV), alkaline-earth metals (\CIII),  $n p^1$ (\CII), $n p^2$ (\CI, \NII, \OIII, \SIII), $n p^3$ (\NI, \SII, \NeIV) and $n p^4$ (\OI)  were chosen for the analysis. Also, \OIIIp, \SIIp, \NIIp\ and \OIp\ lines are the most common to research extended objects. The list of lines is presented in Table~\ref{tab:ListOfLines}.

\begin{table}
        \begin{tabular}{lccc}
                \hline
                \hline
                Number & Element    & Ionisation stage & Wavelength of components, \AA    \\
                \hline
                1  & H  & \ion{}{i}   & 6563 \\%
                2  & C  & \ion{}{i}   & 4622, 4627 \\%
                3  &    & \ion{}{ii}  & 4737, 4739, 4746, 4749 \\%
                4  &    &     & 6580 \\%
                5  &    & \ion{}{iv}  & 5803, 5814 \\%
                6  & N  & \ion{}{i}   & 5198, 5200 \\%
                7  &    & \ion{}{ii}  & 6527, 6548, 6583 \\%
                8  &    &     & 5755 \\%
                9  & O  & \ion{}{i}   & 6300, 6364, 6394 \\%
                10 &    & \ion{}{iii} & 4363 \\%
                11 &    &     & 4933, 4959, 5007 \\%
                12 & S  & \ion{}{ii}  & 6716, 6731 \\%
                13 &    &     & 4069, 4076 \\%
                14 &    & \ion{}{iii} & 6312 \\%
                15 & Ne & \ion{}{iv}  & 4714, 4717, 4724, 4726 \\%
                \hline
        \end{tabular}
        \caption{List of lines and multiplets, in which intensity maps are calculated in this work.}
        \label{tab:ListOfLines}
\end{table}

We upgraded a model of \CIV\ (to 24 levels). We used effective collision strengths from \cite{2004PhyS...69..385A} and radiative transitions data from Chianti v10.1 \citep{2021ApJ...909...38D}. Extension of the model allowed us to study more spectral lines. In the future, extension of H and He atoms and ion models will be useful.

\subsection{Calculated models}
\label{s:res_m}

\begin{figure}
        \includegraphics[width=\textwidth]{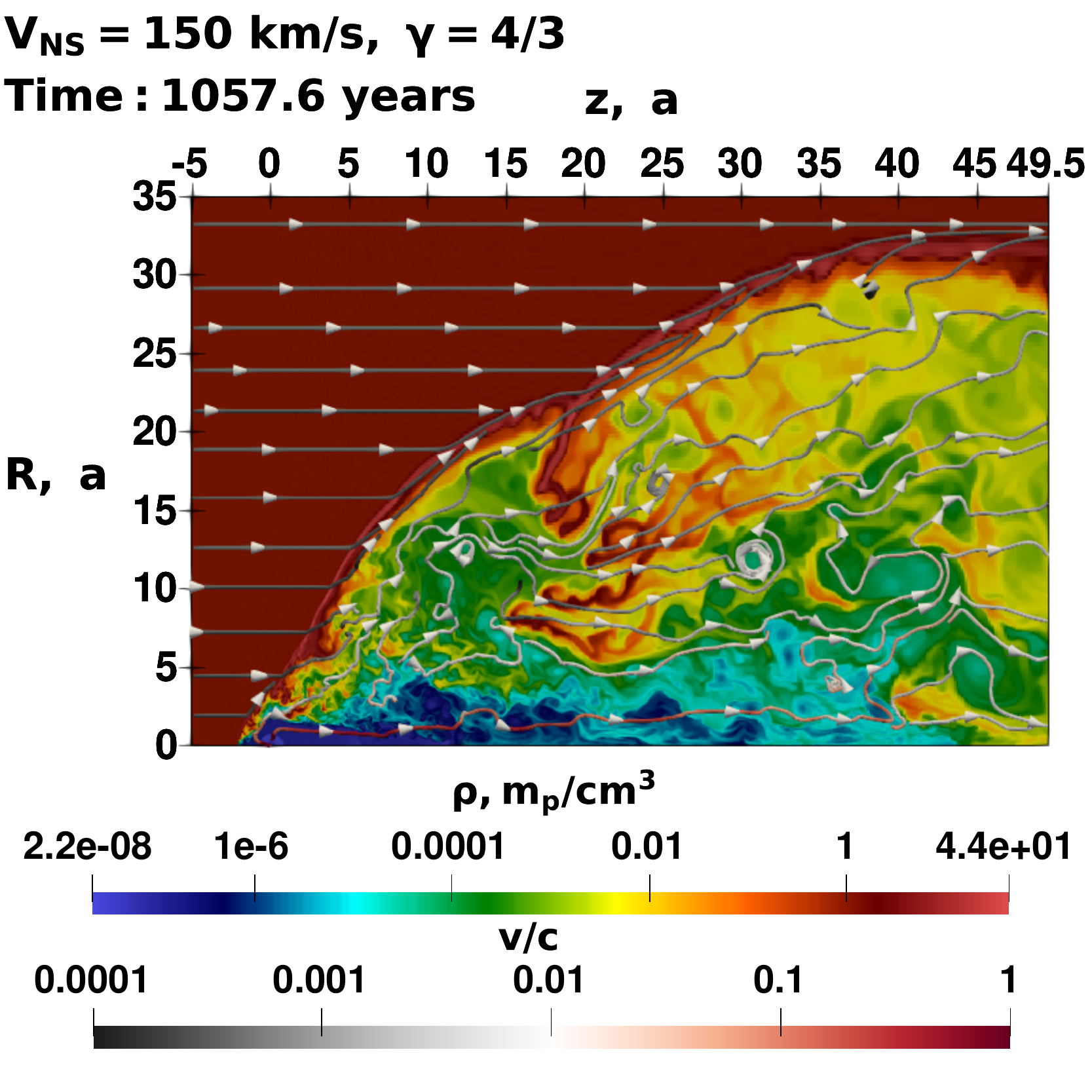} 
        \caption{Distribution of density and velocity in pulsar frame of reference in v01g43nv model. The logarithm of density in $m_p / \text{cm}^3$ is shown by colour. The velocity field is shown with streamlines with arrows.}
        \label{fig:rhoV1}
\end{figure}

\begin{figure*}
        \includegraphics[width=\textwidth]{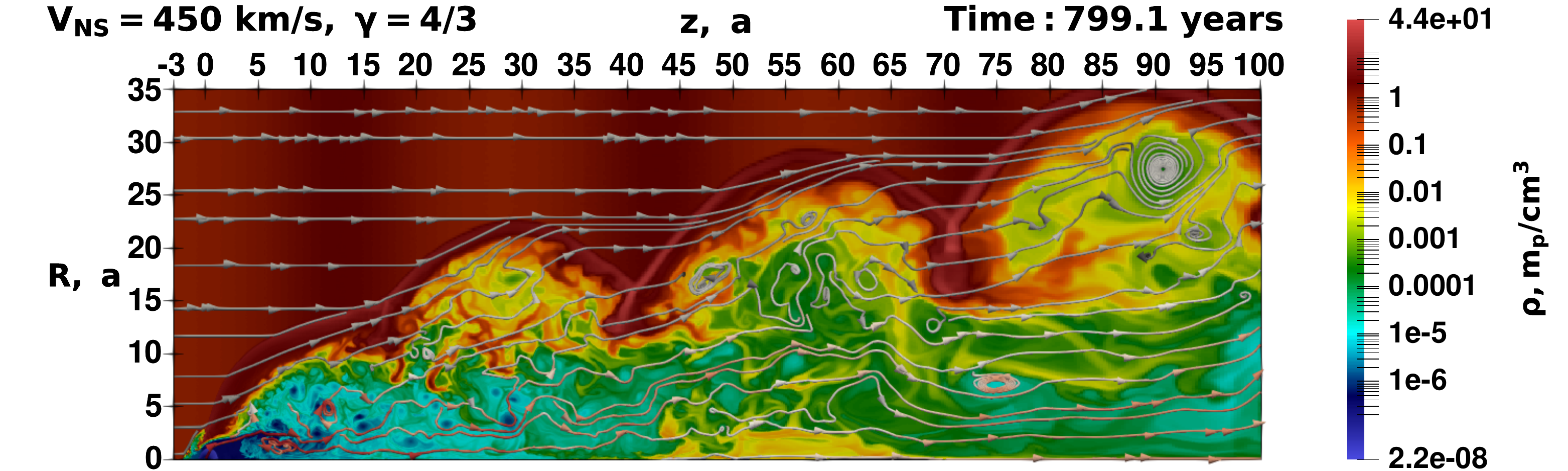}
        \includegraphics[width=\textwidth]{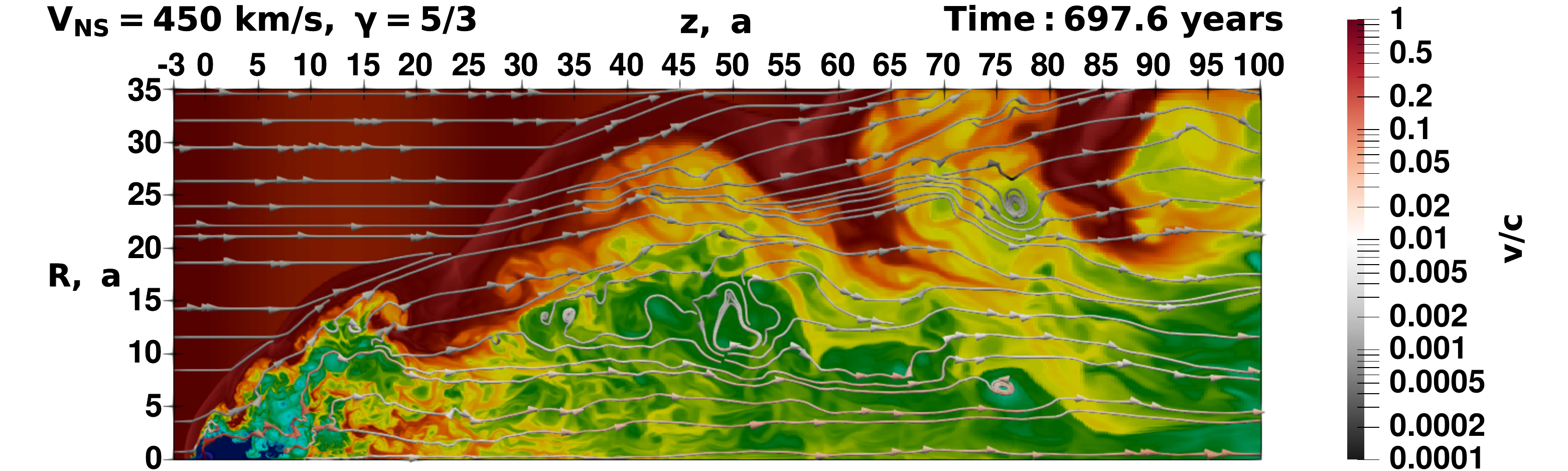}
        \includegraphics[width=\textwidth]{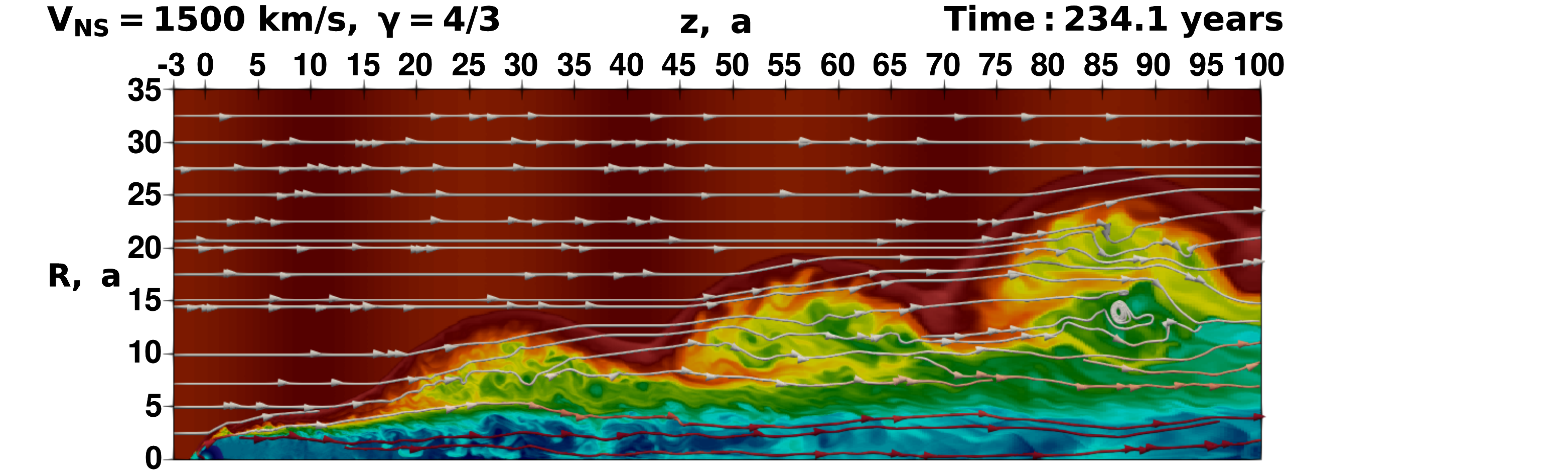}
        \includegraphics[width=\textwidth]{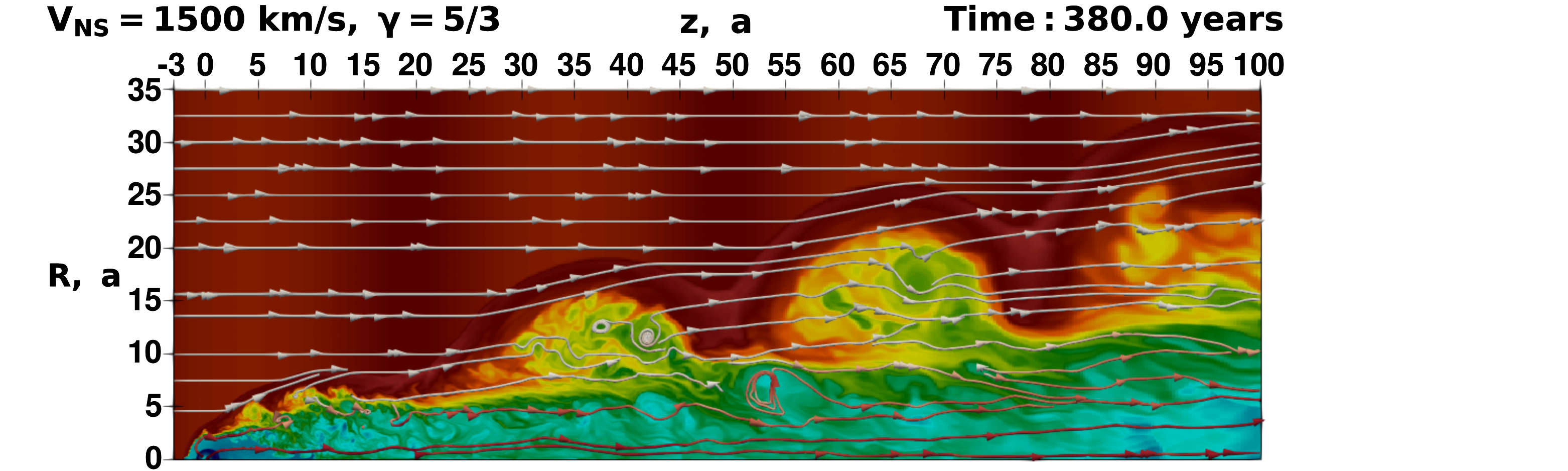}
        \caption{Same, as Figure~\ref{fig:rhoV1}, for v03g43, v03g53, v1g43 and v1g53  models.}
        \label{fig:rhoV2}
\end{figure*}

In every model, we obtained an equilibrium quasi-stationary solution. Density maps and flow streamlines are shown in Figures~\ref{fig:rhoV1} and~\ref{fig:rhoV2}.

There is a forward shock on the outer part of the nebula (crimson colour on Figures~\ref{fig:rhoV1} and~\ref{fig:rhoV2}), on which neutral ISM is shocked and compressed. A zone of shocked ISM is spanning inwards until a contact discontinuity (mostly yellow-green colour), which is present on the head of the nebula at $V_{NS} = 450$~km/s. Then due to strong mixing it breaks down via Kelvin-Helmholtz instability. At $V_{NS} = 1500$~km/s a contact discontinuity remains intact not only on the head, but on a following bubble, even though with some mixing. In the tail of the nebula there is a vast zone of shocked pulsar wind mixed with ISM (yellow and green colours). At the inner part of the shocked wind zone, on the boundary with an unshocked wind (dark blue colour), a reverse shock with a Mach disk is formed (a jump in density, depicted as jump in colour). It is located near the position of the pulsar: $(R,~z) = (0,~0)$.

The main factor impacting the nebula morphology is the pulsar velocity. For the value $450$~km/s the interaction with the perturbations in the ISM is relatively long and distinctive, so-called bubbles are forming in the tail of the nebula. When the pulsar passes a region with low ISM density, the bow shock starts almost isotropic expansion and forms a close to spherical bubble in the tale of the nebula. In the high density region, the shock wave propagates slower, forming a so-called neck zone. At high pulsar velocity $1500$~km/s, bubbles are characterised by a notably smaller size. In the high velocity case, we see less prominent mixing of shocked ISM matter with pulsar wind one. 

Another kind of features presented in our models and also observed by \cite{2014ApJ...784..154B} in some objects is formation of so-called rings, where shocks of two bubbles collide. In the area of the collision, the density of the matter grows rapidly and can reach the value $\sim$80~$m_p/$cm$^3$, and the number density of electrons can be as high as $\sim$40~$/$cm$^3$. The temperature is not high (falls to a hundred thousands K) as in other shocked regions. Taking into account low matter density and short dynamical timescale, these conditions favour a low ionisation stages of atoms.

The solution varies noticeably with the adiabatic index ($\gamma$) of the ideal gas. The distinction between values is in the compression ratio in strong shocks, which is equal to 7 for $\gamma$~=~4/3 and 4 for $\gamma$~=~5/3. The first case is the ultrarelativistic limit for the adiabatic index, which is applicable to strongly relativistic flows. However, it results in density behind shocks being overestimated by almost a factor of 2. The last case is the classical limit that is applicable to shocks in the ISM. But relativistic winds' ability to compress becomes underestimated, that leads to intensive mixing with ISM. Thus, the overall morphology of the nebula is better described in case of $\gamma$~=~4/3, but shocks in ISM, which are essential for calculating the emissivity, are more realistic in case of $\gamma$~=~5/3. In reality, one can expect behaviour in between of these cases.

Direct comparison of flow morphology in the low speed model ($V_{NS}=$~150~km/s) with high velocity models is not straightforward due to initial uniform density of ISM, as shown in Table~\ref{tab:models}. Also, due to low pulsar speed and fast expansion of the bow shock in radial direction, the morphology of the nebula in this case only developed in $z<$~50~a region. Further simulation was meaningless; Mach cone leaves the computation domain. Nevertheless, we are able to highlight some distinct features of the model. The interaction of pulsar wind with ISM is the most active and strong. Due to Kelvin-Helmholtz instability, ISM matter actively mixes with pulsar wind, forming a complex and dynamic inner structure of nebula.

\section{Synthetic intensity maps}\label{sec:maps}

In the case of intensity maps appears a new free parameter $\chi$ -- angle between $V_{NS}$ and picture plane. How intensity maps react to the variation of $\chi$ was investigated in the paper \cite{2020MNRAS.497.2605B}. Here we calculated intensity maps of fast moving pulsar nebulae in various spectral lines for $\chi=0.2$~rad. We present mapped values of intensity, which are unchanged with varying distance to the object if the extinction is not significant. Angular size of nebula can vary, so we plot intensity against physical size. The coordinates of intensity maps are $(X, Y)$, their plane is rotated by $\chi$ around $R$ (or $X$) axis with respect to $(R, z)$, similar to \cite{2020MNRAS.497.2605B}.

\begin{figure*}
        \includegraphics[height=0.44\textheight,angle=-0, trim=0 0 0 0]{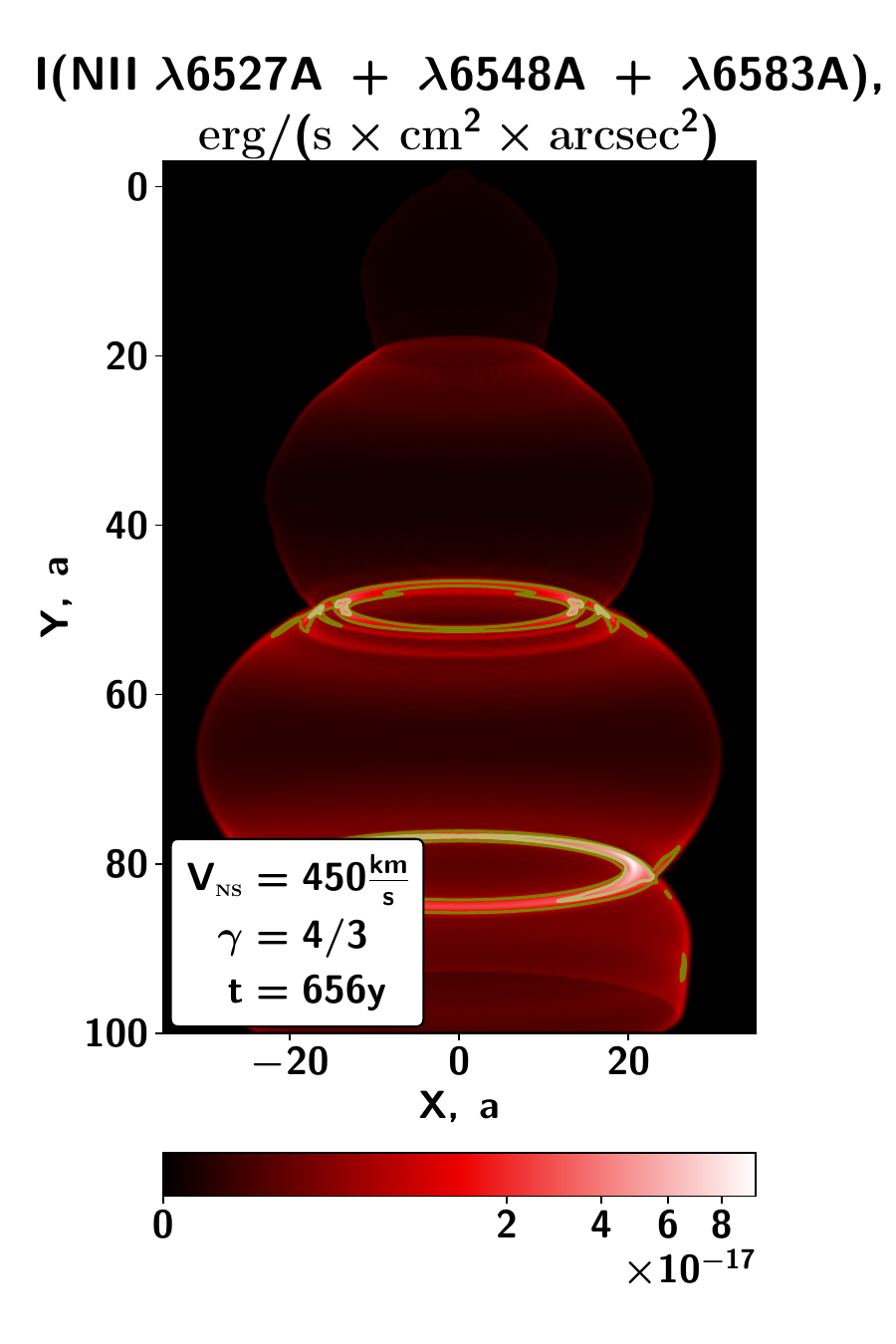}
        \includegraphics[height=0.44\textheight,angle=-0, trim=0 0 0 0]{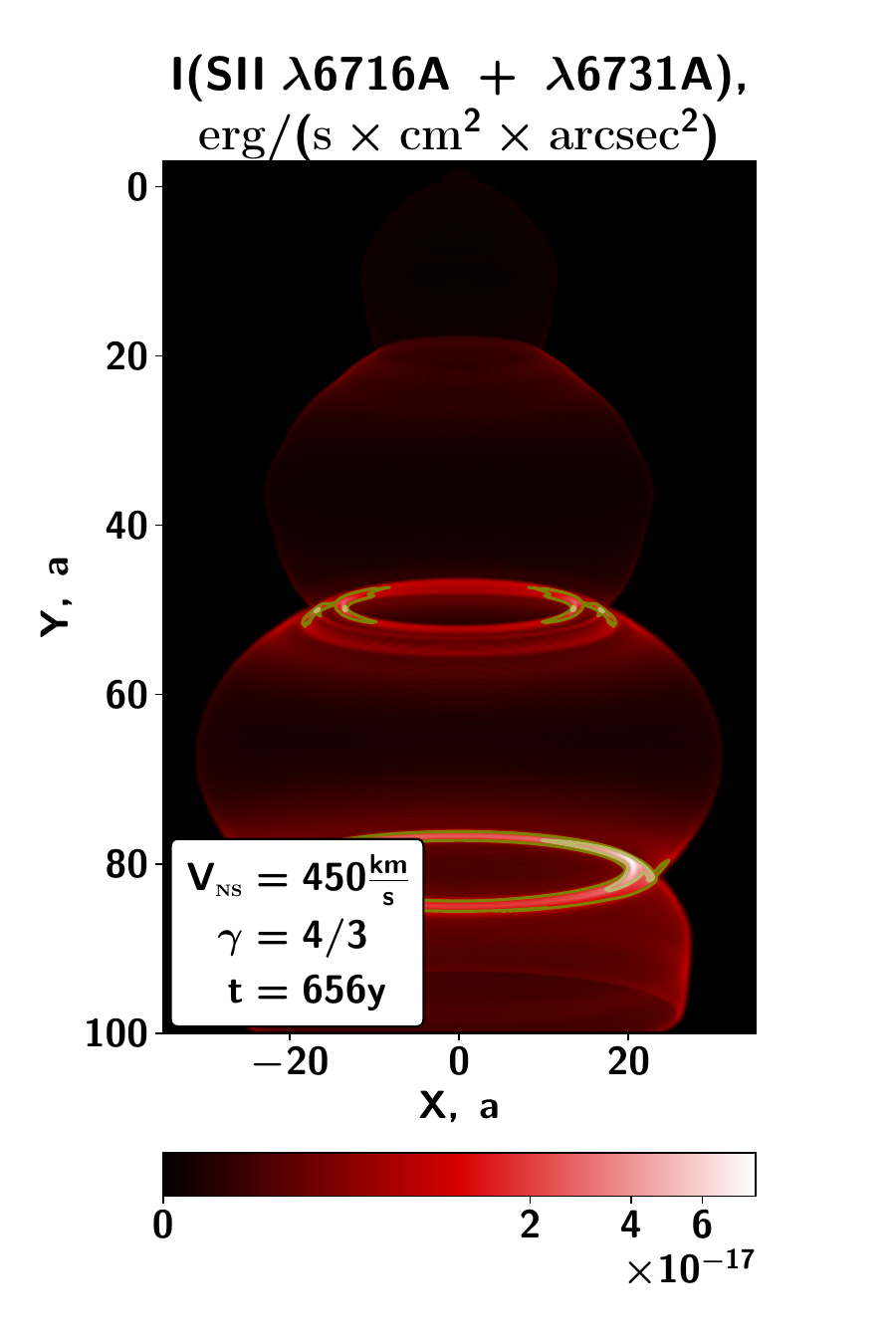}
        \caption{Synthetic intensity maps in the brightest optical lines of singly ionised atoms in calculation: \NIItriplet\ (left panel), \SIIdoubletMain\ (right panel).  Contours highlight 1, 3 and 10 $I_{17}$.}
        \label{fig:LinesSinglyIon}
\end{figure*}

\begin{figure}
        \includegraphics[height=0.44\textheight,angle=-0, trim=0 0 0 0]{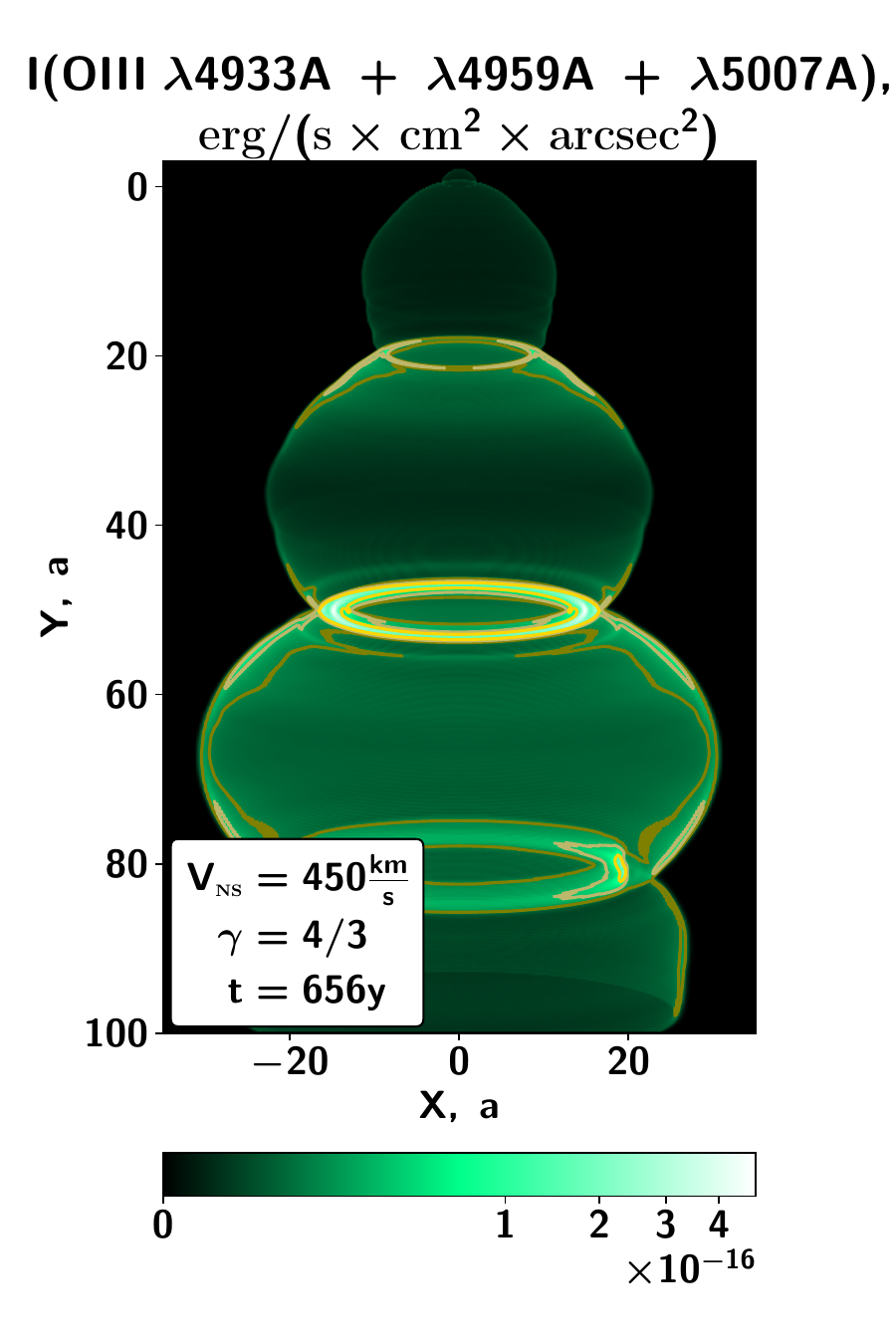}
        \caption{Synthetic intensity maps in the brightest optical lines of doubly ionised atoms in calculation: \OIIItriplet.  Contours highlight 1, 3 and 10 $I_{17}$.}
        \label{fig:LinesDoublyIon}
\end{figure}

The maps are presented in Figures~\ref{fig:Ha1}~--~\ref{fig:LinesDoublyIon} and share some common features. Firstly, emitting regions has a layered structure, with a layer having a surface and near-certain depth (smoothness of emitting layers on presented maps is partly due to 2D calculations).

By the reason of projection effect, nebulae are much brighter to the edges of bubbles -- a bulk of emitting material lies on the line of sight there. This is shown by contours on intensity maps, which represent typical values for detection of extended emission on relatively low signal level (we assume a typical value of $10^{-17}\frac{\text{erg}}{\text{s} \times \text{cm}^2 \times \text{arcsec}^2}$, olive colour); higher signal levels $3\times10^{-17}$ and $10^{-16}\frac{\text{erg}}{\text{s} \times \text{cm}^2 \times \text{arcsec}^2}$ are indicated by dark khaki and gold colours, respectively. Those values are shared by the contours in all the figures presented.

\subsection{Emission maps in lines of neutral atoms}

\begin{figure}
        \includegraphics[width=0.99\textwidth,angle=-0, trim=0 0 0 0]{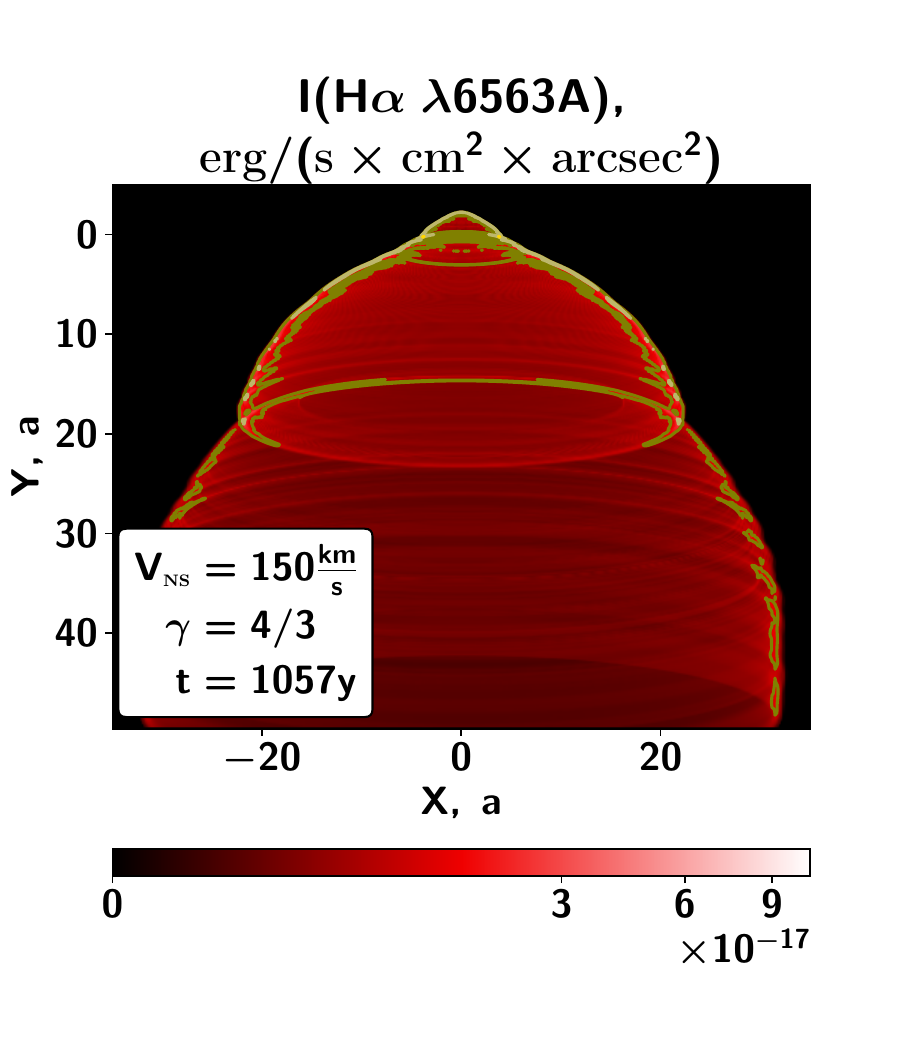}
        \caption{Synthetic intensity map of model v01g43nv nebula in \Ha. The contours highlight levels 1 (olive colour), 3 (dark khaki) and 10 $I_{17}$ (gold).}
        \label{fig:Ha1}
\end{figure}

\begin{figure*}
        \includegraphics[height=0.45\textheight,angle=-0, trim=0 0 0 0]{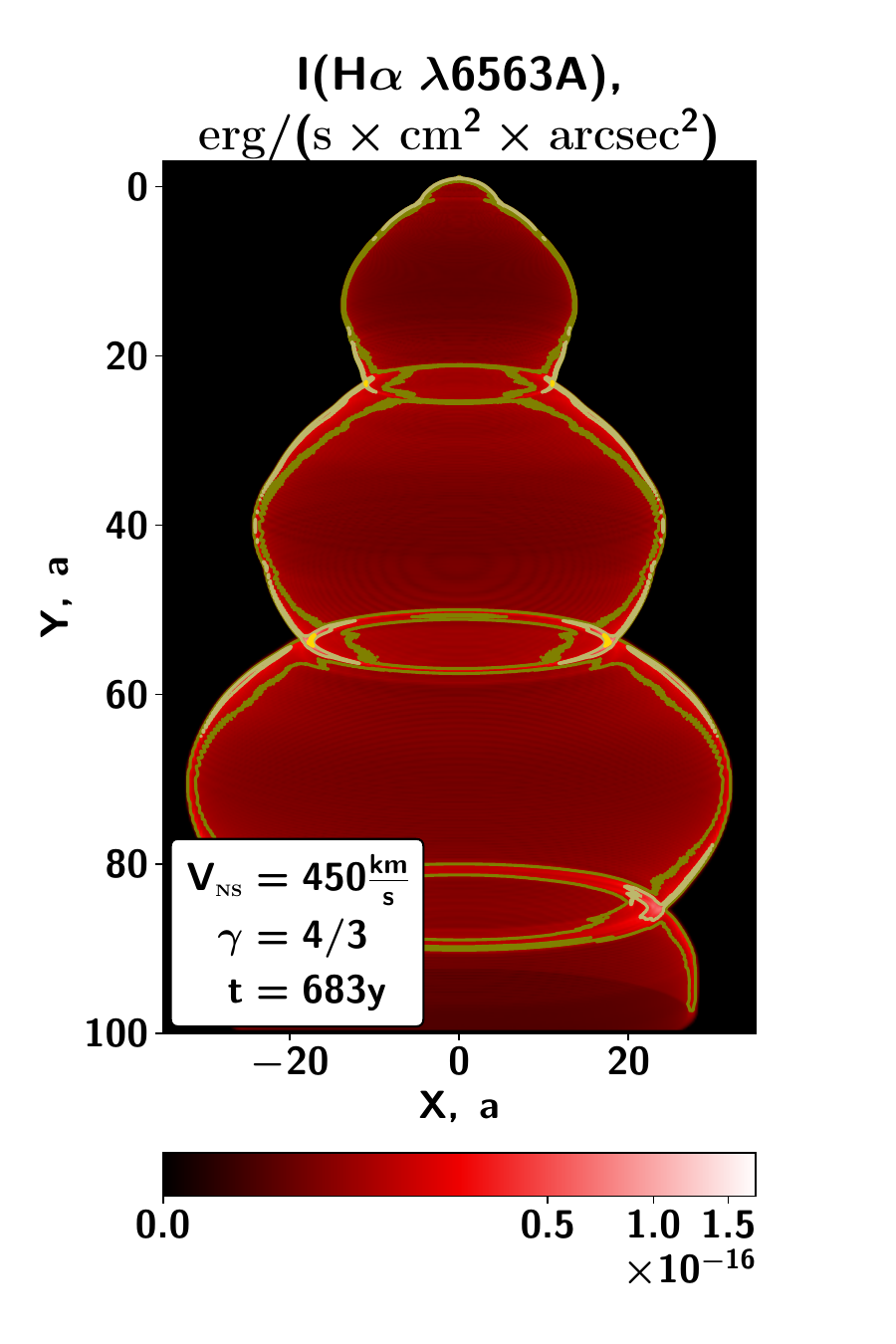}
        \includegraphics[height=0.45\textheight,angle=-0, trim=0 0 0 0]{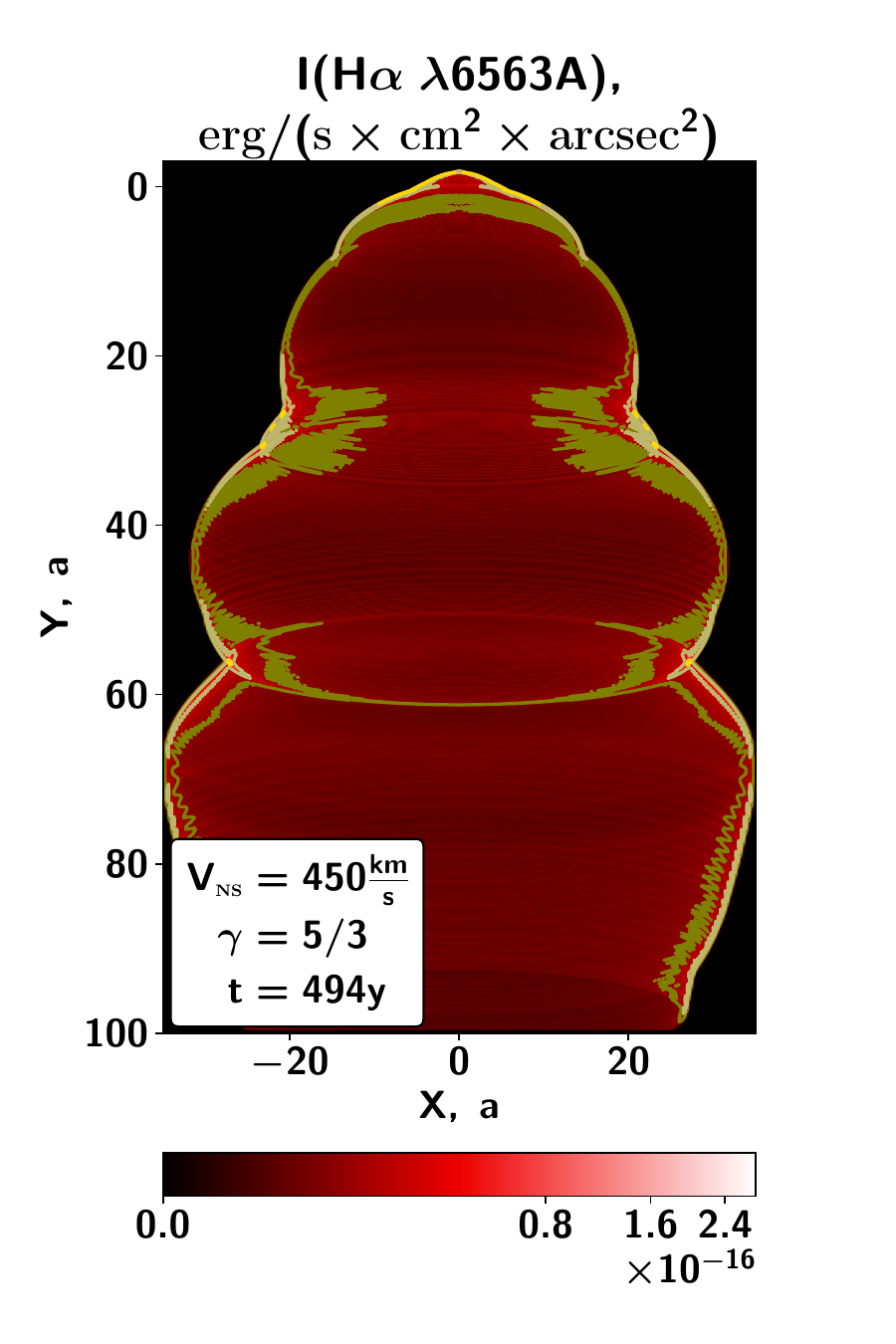}
        \includegraphics[height=0.45\textheight,angle=-0, trim=0 0 0 0]{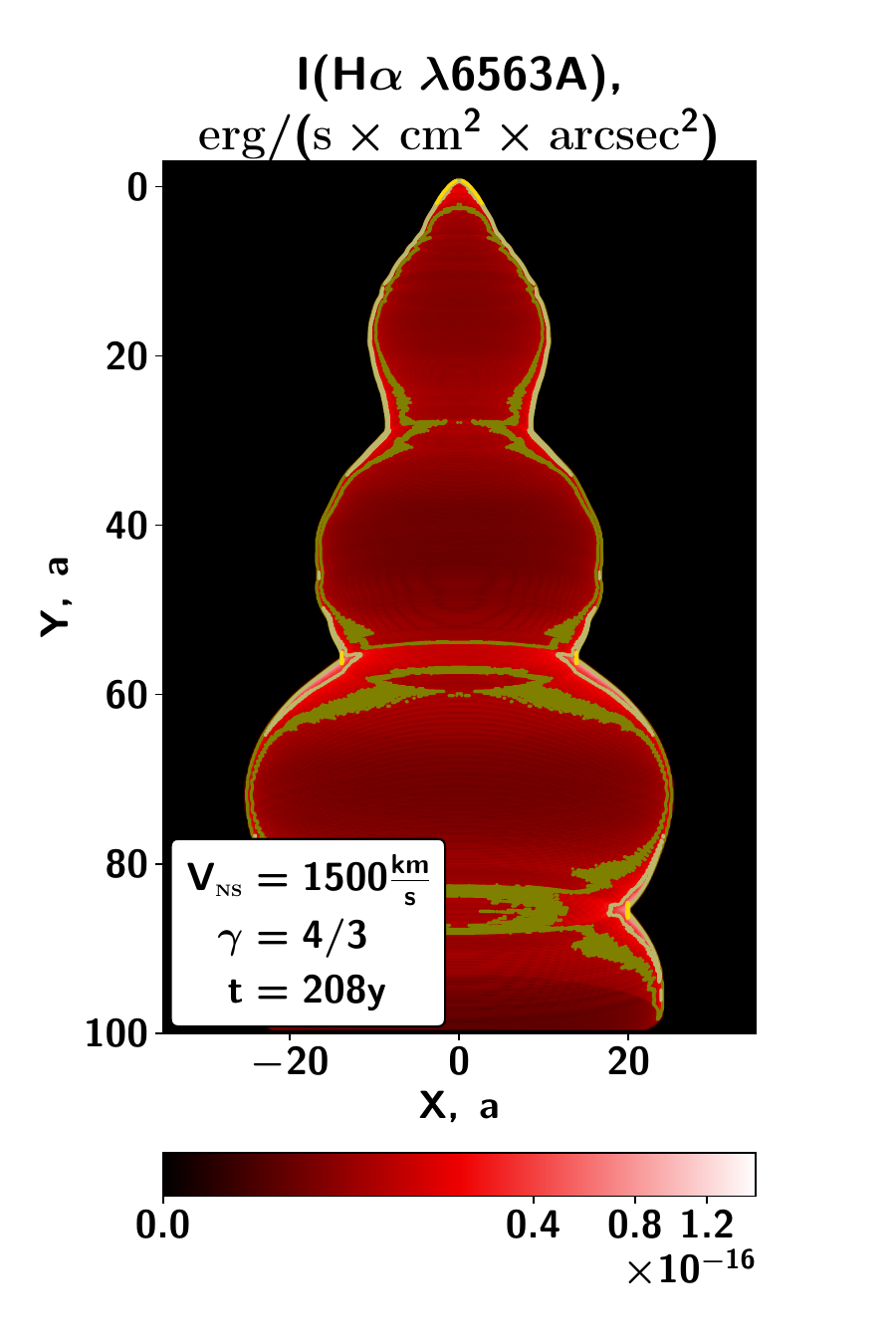}
        \includegraphics[height=0.45\textheight,angle=-0, trim=0 0 0 0]{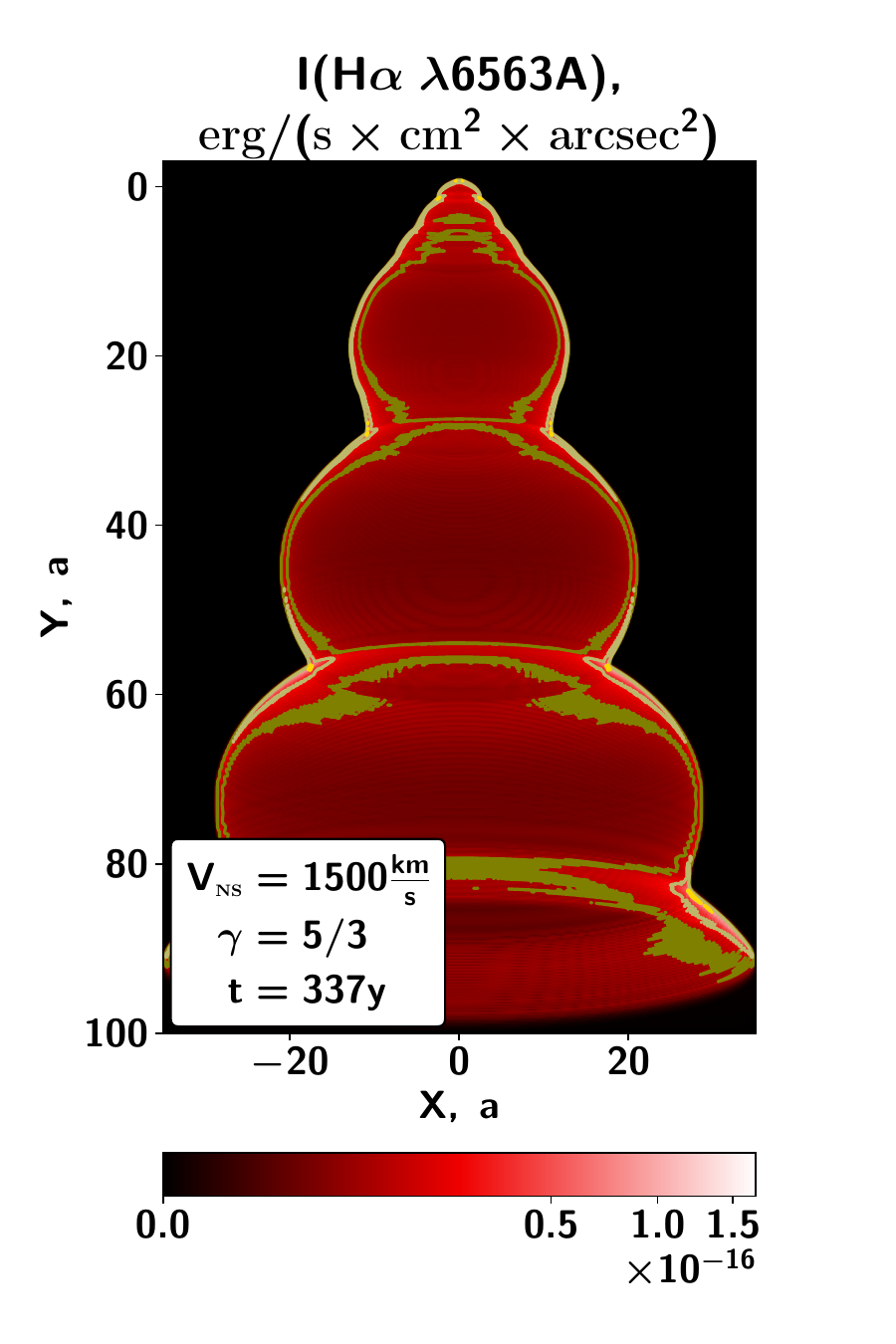}
        \caption{Same, as Figure~\ref{fig:Ha1}, for v03g43 (top left panel), v03g53 (top right panel), v1g43 (bottom left panel) and v1g53 (bottom right panel) models.}
        \label{fig:Ha2}
\end{figure*}

\Ha\ is the most important line for research of bow-shock PWNe. Now, it is the only line in which these objects are systematically observed and can be directly compared with numerical models \citep[see ][]{2020MNRAS.497.2605B}. Furthermore, it is quite useful as a standard for analysis of other spectral lines and a marker of modern observational possibilities. Therefore, we calculated intensity maps for it first. They are presented on Figures~\ref{fig:Ha1} and~\ref{fig:Ha2}. We also performed comparison with observations, which is described in Section~\ref{subsec:ANTFcomparison}.

Only a thin layer of ISM matter just behind the bow shock emits in \Ha, since \HI\ exists only in low temperature plasma and maximum emissivity is achieved around $X$(\HI)~$\approx$~0.5. It is being reached by collisions with energetic electrons in short after ISM matter passes the bow shock ($\sim 5$ cells). On the one hand, this feature allows an easy detection with observations and a direct study of the environment near bow shock. But on the other, \Ha\ photons are not being emitted from deeper regions of the nebula. It makes gathering information about deep volume structure in this waveband impossible, and observations in other lines become highly demanded.


\begin{figure*}
        \includegraphics[height=0.45\textheight,angle=-0, trim=0 0 0 0]{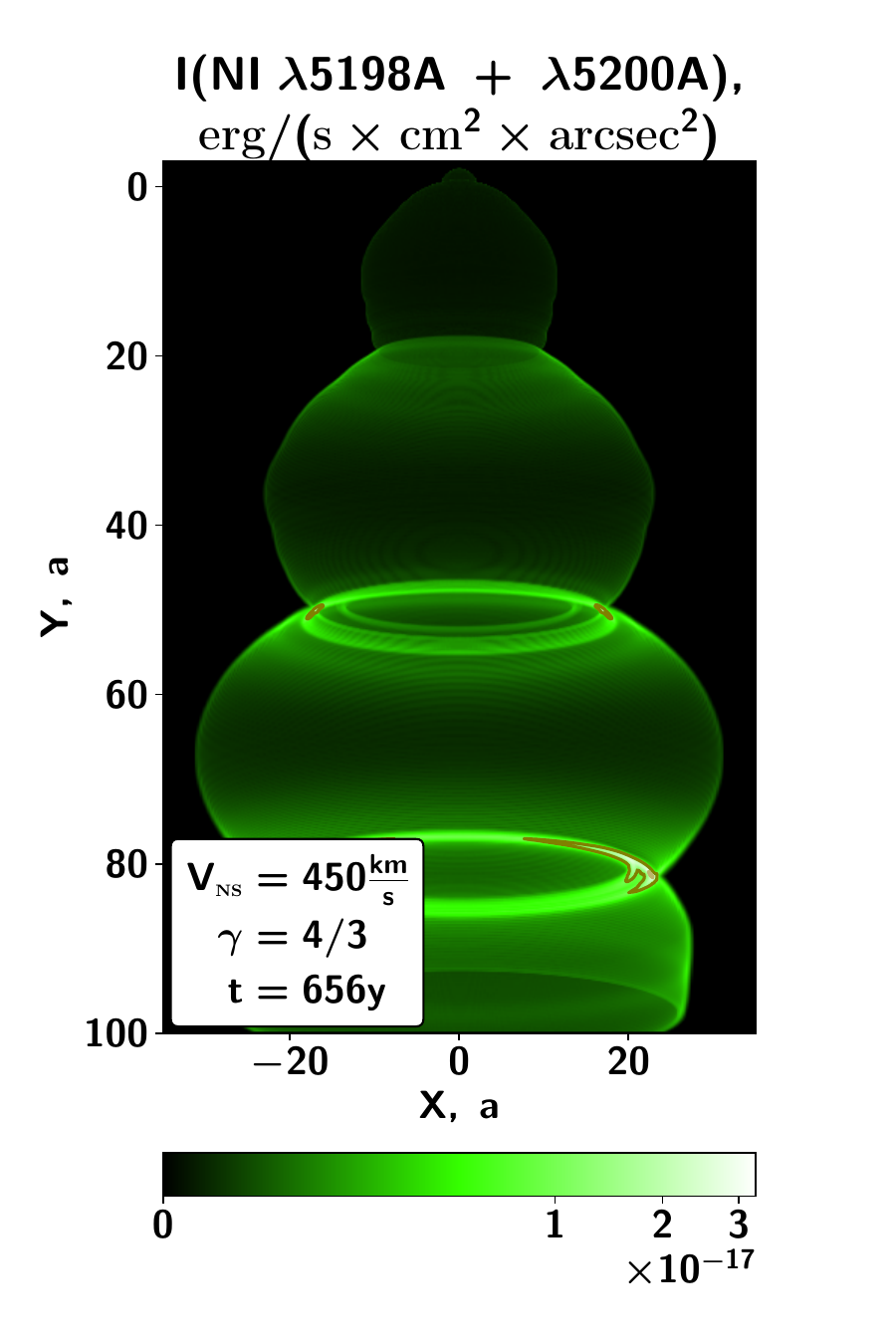}
        \includegraphics[height=0.45\textheight,angle=-0, trim=0 0 0 0]{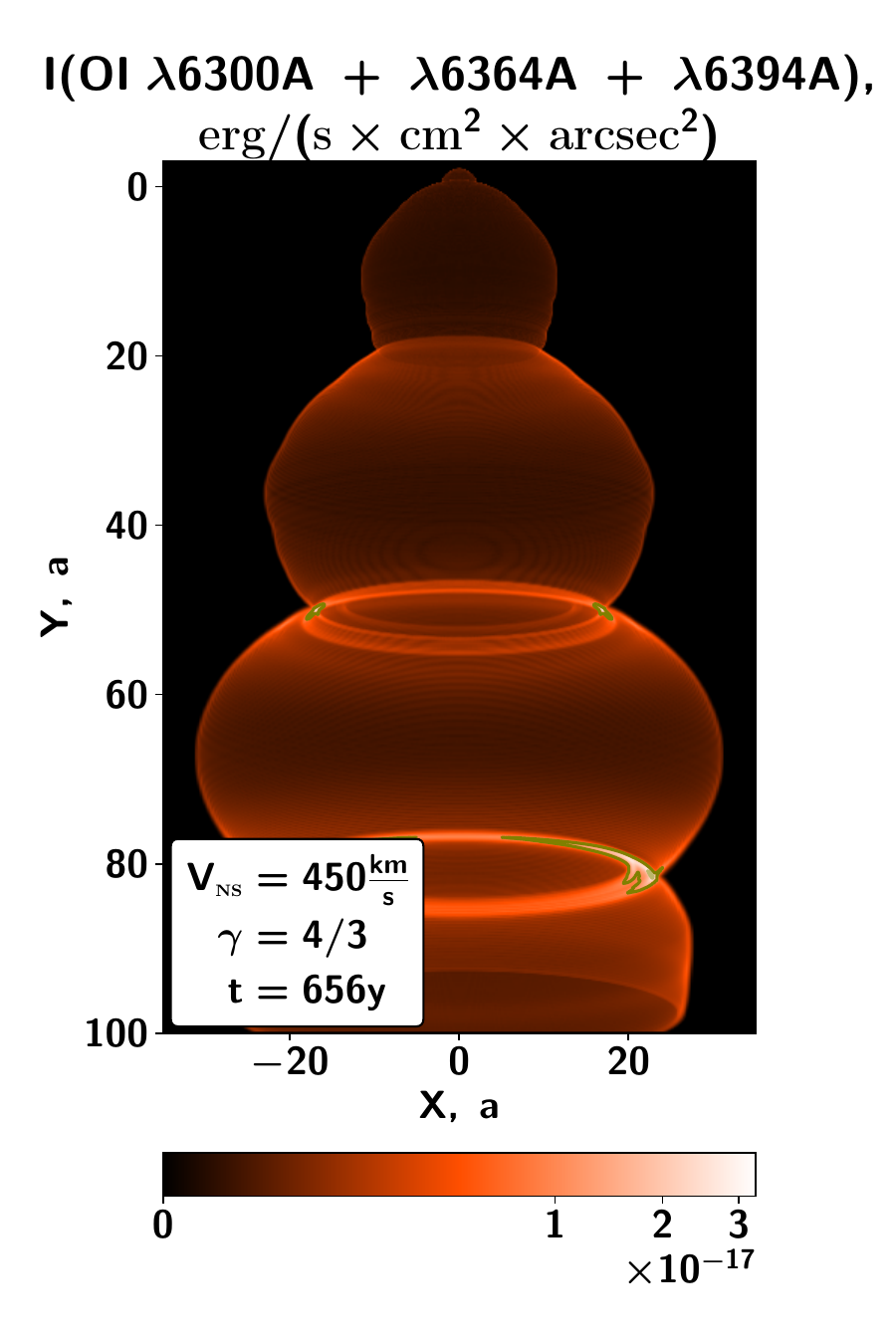}
        \caption{Synthetic intensity maps in the brightest optical lines of neutral atoms in calculation: \NIdoublet\ (left panel), \OItriplet\ (right panel). Contours highlight 1, 3 and 10 $I_{17}$.}
        \label{fig:LinesNeutral}
\end{figure*}

Other bright lines of neutral atoms are \NIdoublet\ doublet and \OItriplet\ triplet with the 6300~\AA\ brightest component (synthetic intensity maps are presented in Figure~\ref{fig:LinesNeutral}). These lines mostly highlight the same features as \Ha\ and highlight the bow shock of the nebula. We see a less uniform intensity distribution, with slight humps on the front and back parts of the bubbles. The other difference is that, unlike \Ha, centres of bubbles have lower intensity and rings become visible. 

\subsection{Emission maps in lines of singly ionised atoms}

The brightest lines of singly ionised atoms are \NIItriplet\ nebular triplet and \SIIdoubletMain\ doublet. Synthetic intensity maps of them are shown in Figure~\ref{fig:LinesSinglyIon}. Spectral lines of atoms in the second ionisation stage are placed on the intermediate depth relatively to shock front. Emission comes from a thin layer at bow shock, the head of nebula is dimmer and rings are brighter than in the case of neutral atoms. Regions of rings stand out due to high density and relatively low temperature. Thickness of the rings are $\sim 2a$ with $T<40000$~K in v03g43 model, density reaches $\gtrsim 10~m_p/\text{cm}^3$. 

In v03 models, regions of line formation start shifting to rings, although their inner structure is only slightly noticeable. Bow shock is strongly expressed. In v01g43nv model the same features appear, but bright filaments are formed from ISM matter as a product of turbulence in shocked pulsar wind.  

\subsection{Emission maps in lines of doubly ionised atoms}

The \OIIItriplet\ triplet is the most bright lines of twice ionised atoms. Synthetic intensity maps of it are shown in Figure~\ref{fig:LinesDoublyIon}. \OIII\ ions and thus their emission exists in a thick layer of a favourable temperature regime. They form rings, which are more pronounced than rings in other lines. Its emission zone lays deep behind the bow shock on the intensity maps. The cross-section radius of nebula bubbles is smaller than in \Ha. Features mentioned above make observations and detection in \OIIIp\ easier due to high contrast. The diffuse background emission of ISM is smaller in the case of \OIIIp\ triplet than in \Ha. 

\section{Comparison between observations and models for \Ha}\label{sec:Ha_comparison}
\subsection{Scaling flux of observed objects using ATNF radio data}\label{subsec:ANTFcomparison}

\begin{figure*}
        \includegraphics[width=\textwidth, angle=-0, trim=0 0 0 0]{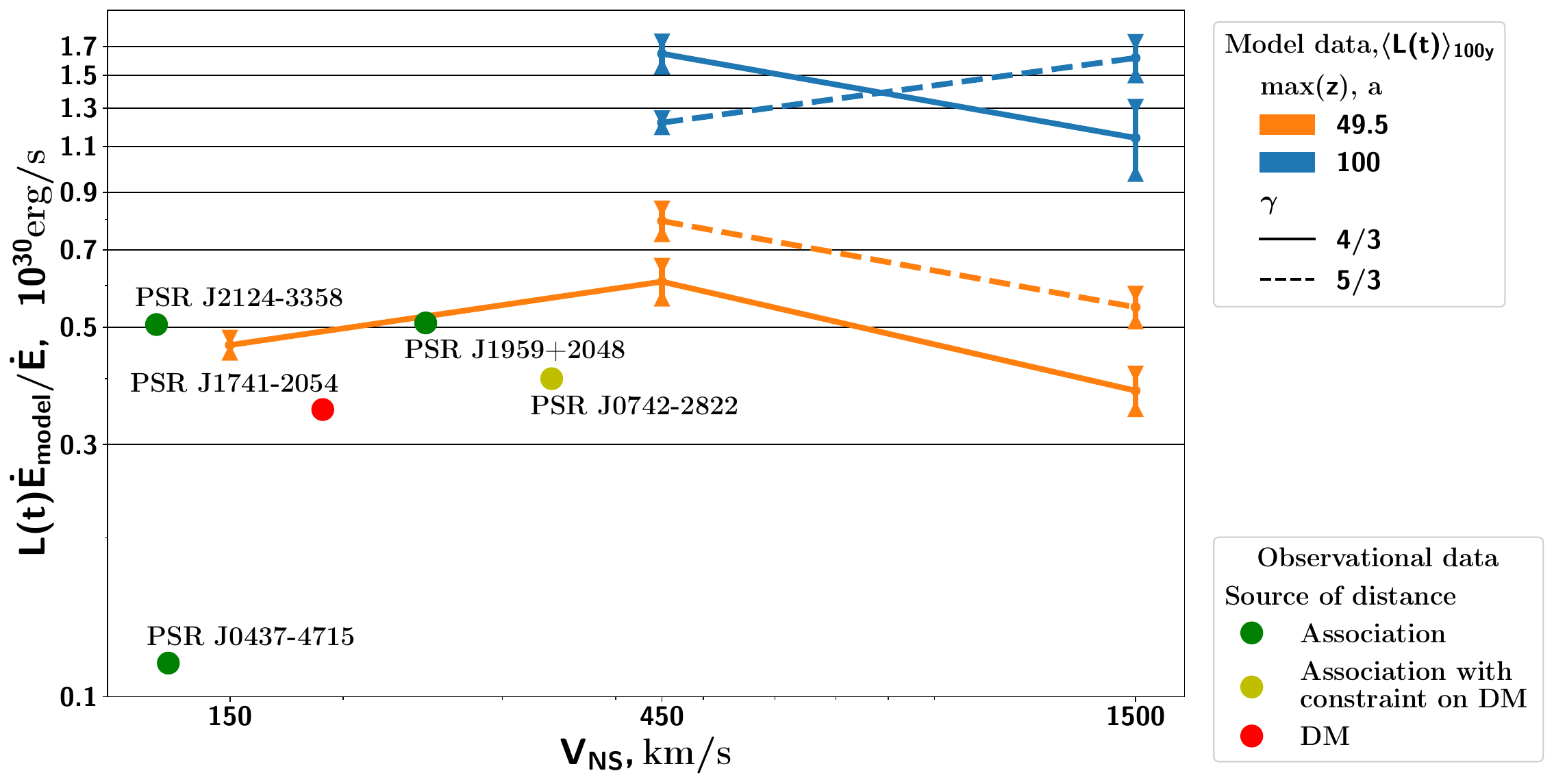}
        \caption{\Ha\ luminosity of bow-shock PWNe versus pulsar velocity. Observed nebulae were calibrated using Equation~(\ref{eq:calibration}). For model nebulae, luminosity averaged over the last 100~years of simulation is presented. Error bars show standard deviation over the mentioned period. Remark: DM -- dispersion measure.}
        \label{fig:Ha_mod_vs_obs_comp_luminosity}
\end{figure*}

For verification of our results, we compare integrated flux of the models and observed flux in \Ha\ line for 7 nebulae. We take into account integrated fluxes, extinction in R-band \cite{2014ApJ...784..154B}, spin-down power, proper motion velocity and distances from Australia Telescope National Facility catalogue version 1.70 (ATNF, see \cite{2005AJ....129.1993M}).

We calculate PWNe luminosity as:
\begin{equation}
        L _{\text{H}\alpha} = F_{T,~\text{H}\alpha} \frac{h c}{\lambda_{\text{H}\alpha}} 4 \pi D^2 10^{0.4 A_R},
\end{equation}
where $F_{T,~\text{H}\alpha}$ -- observed integrated \Ha\ flux in cm$^{-2}$s$^{-1}$, $h$ -- Plank's constant,  $c$ -- speed of light, $\lambda_{\text{H}\alpha}$ -- \Ha\ wavelength (6563~\AA), $D$ -- distance to pulsar, $A_R$ -- extinction in red waveband.

For most PWNe, the pulsar speed is measured only in the plane of the sky ($V_\tau$). Assuming a general population of pulsars' velocity is being isotropically distributed in space, and their local ISM is at rest in respect to the Galactic rotation \citep[][give detailed description of determining $V_\tau$]{2005AJ....129.1993M}, we can calculate spatial velocities relative to the local ISM as:
\begin{equation}\label{eq:velocities}
        V_{NS} = \sqrt{3/2} V_\tau.
\end{equation}

The parameters of bow-shock PWNe, which were defined from optical observations \citep{2014ApJ...784..154B}, strongly correlate with each other (for example, the flux from the nebula's head and its size are practically linearly dependent on log -- log scale). It could be caused by natural reasons as selection effects of observations. At the moment, we cannot distinguish the true nature of the observed correlation. We use only data from radio wavelengths in order to calibrate luminosity. Independent data sources are highly demanded. 

Properties of the observed nebulae vary and don't coincide with models' parameters. This fact requires the development of a calibration procedure, which can be applied to the bow-shock PWNe. In the Appendix~\ref{sec:cal} we discussed various procedures of the calibration. Finally, we attained an equation for \Ha\ luminosity as:
\begin{equation}\label{eq:calibrationtxt}
        L_{\text{H}\alpha}^{calibrated} (V_{NS}) = L_{\text{H}\alpha} \dot{E}_{model}(V_{NS}) / \dot{E}.
\end{equation}
Here $\dot{E}_{model}$ -- spin-down power of model pulsar. This quantity corresponds to model $\rho_{ISM} = 1~m_p/\text{cm}^3$ and $r_s = 0.57$~a, and given $V_{NS}$, according to Equation~(\ref{eq:rs_short}).

In Figure~\ref{fig:Ha_mod_vs_obs_comp_luminosity} we present the comparison of \Ha\ luminosity versus pulsar velocity between model and observed nebulae. Most values are close to models with $\gamma=4/3$ and $max(z) = 49.5$~a. The models are consistent with the observational data. The consistency is present throughout different methods of measuring the distance to pulsars, which strongly affects the luminosity estimation.

Nebula of PSR~J0437-4715 shows 0.5 orders less luminous than others. It has the largest stand-off angular distance (9'') of all observed nebulae, so we presume that lacking luminosity may belong to regions outside the telescopes' field of view. There may be the case analogous to nebula of PSR~J0742-2822, described by \cite{2014ApJ...784..154B}, when past observations detected nebula's front part with high surface brightness, and following ones discovered a tail having lower surface brightness, but due to its size giving noticeable contribution to overall nebula's flux in \Ha. We also cannot fully exclude the hypothesis of pre-ionisation by non-thermal emission, due to which ISM material is partially ionised before passing bow shock. The possibility of this scenario was discussed in \cite{1993Natur.362..133C, 2002ApJ...575..407C, 2014ApJ...784..154B}.

Nebulae of PSR~J1856-3754 and PSR~J2225+6535 were excluded from the comparison. For the first one, there is a possibility of it being the photoionisation nebula, which was discussed by \cite{2001A&A...380..221V}. In the second case, ~\cite{2002ApJ...575..407C} measured atypically low stand-off angle ($\sim 0.1$''). Explanation of its value is challenging, especially considering the nebula's long and bright tail. In both cases, further research is required.

Considering dependency of luminosity from pulsar speed, we notice $L_{\text{H}\alpha}$ being approximately constant throughout the entire range of pulsar velocity in the set of models with $max(z) = 49.5$~a, which is closest to observations. In the same time, in our models, the spin-down power is proportional to the pulsar velocity squared. And due to Equation~(\ref{L_VNS}), in order for $L_{\text{H}\alpha}$ to be constant, there must be limitation on $f(V_{NS})$ and its components (see Appendix~\ref{sec:cal} for the details):
\begin{align}
        \dot{E}_{model} &\propto V_{NS}^2, \\
        f(V_{NS}) &\propto V_{NS}^{-2}, \\
        \epsilon_{\text{H}\alpha} \eta^2 &\propto V_{NS}^{-1}.\label{eq:scaling_constraints} 
\end{align}

If $\epsilon_{\text{H}\alpha}$ is approximately independent of the pulsar velocity (Equation~\ref{eq:epsilonRaymond}), then the coefficient $\eta \propto V_{NS}^{-\frac{1}{2}}$. This contradicts \cite{2002ApJ...575..407C}, but seems plausible, because nebula cone shrinks with increase of pulsar velocity.

If $\epsilon_{\text{H}\alpha}$ is proportional to the pulsar velocity in the negative power law, $\eta$ is less defendant in the pulsar velocity. For Equations~(\ref{eq:epsilon_romani1}) and~(\ref{eq:epsilon_romani2}), there is $\eta \propto V_{NS}^{-\frac{1}{4}}$ and $\eta \propto V_{NS}^{-\frac{1}{8}}$ respectively. This also seems plausible and more consistent with \cite{2002ApJ...575..407C}.

\subsection{Morphology comparison}
\label{ssec:morph}

\begin{figure*}
    \centering
    \includegraphics[width=\textwidth]{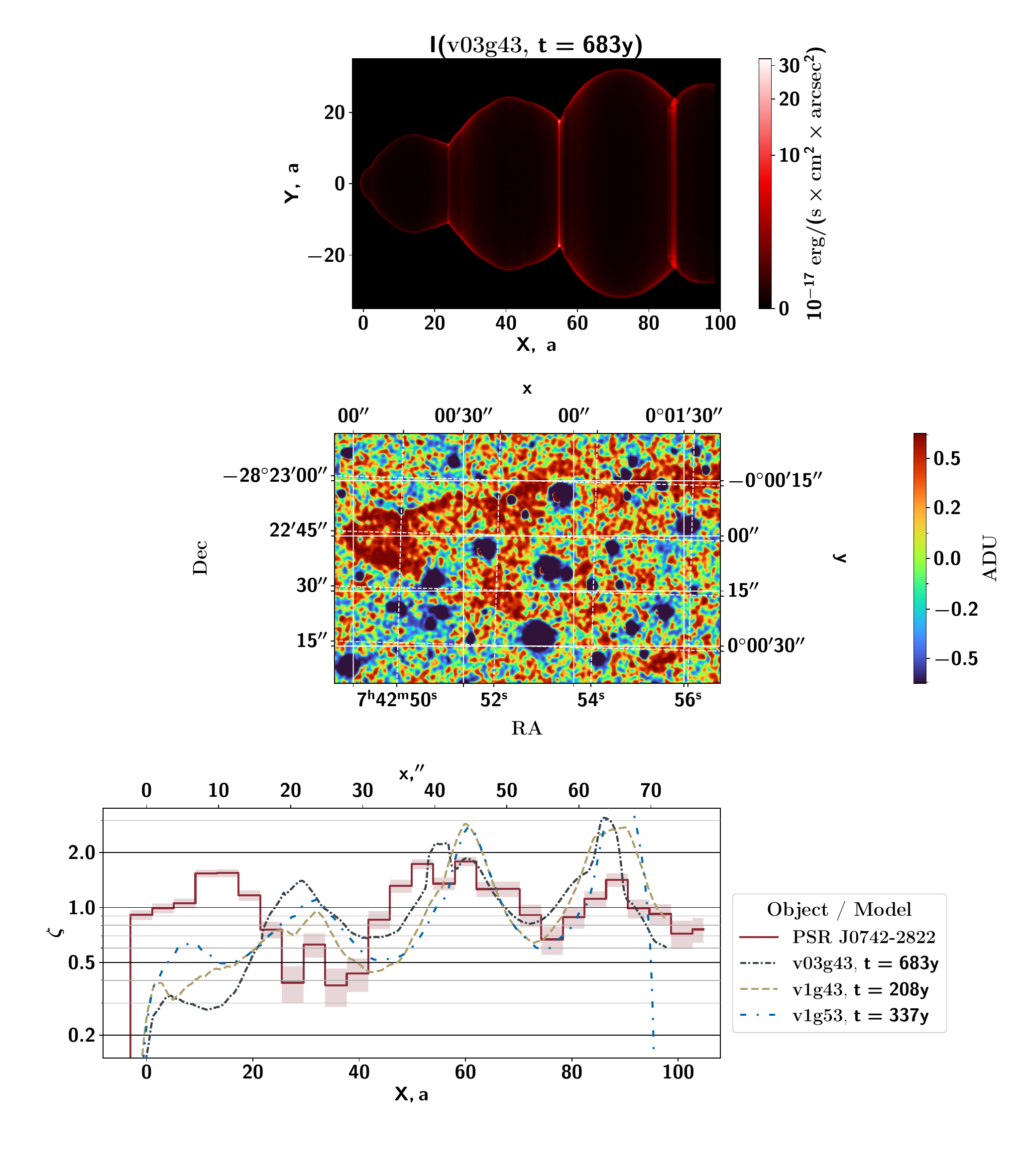}
    \caption{We show \Ha\ intensity map of model v03g43 (motion of pulsar is in picture plane, top panel). Presented by Prof Romani  \citep[][]{2014ApJ...784..154B}, observation data of PSR J0742-2822 nebula in pulsar's frame  of reference smoothed with 0.5'' Gaussian filter (middle panel). Observational binned profile (bin size is 3'') of brightness along tail of nebula and model profiles (smoothed with the size of observational profile bin), bottom panel.}
    \label{fig:morphology_obs}
\end{figure*}

An important criterion of models' correctness is consistency between model and observed nebulae morphology. We compare the nebula PSR J0742-2822 and the model v03g43 profiles in the \Ha\ line. This object was selected due to long tail region and probable proximity of pulsar velocity to picture plane. Observational data was taken from \cite{2014ApJ...784..154B} in analog-to-digital units (ADU) normalised to unknown time interval with already subtracted continuum, so it wasn't possible to get an absolute calibration, with included equatorial World Coordinate System (WCS). 

Movement of the real pulsar is misaligned with the equatorial world system, so to perform a comparison we had to match its direction in the picture plane with models. We also aligned pulsar positions in the sky and in models. Data on pulsar position (J2000) and proper motion were obtained from ATNF v1.70 \citep{2005AJ....129.1993M}. Using {\it astropy} package \citep{2022ApJ...935..167A}, we calculate WCS for the observation epoch that analogous to model in angular units. Then we projected the given frame to new WCS using adaptive resampling with kernel width of 1.3~pixels from {\it astropy} affiliated {\it reproject} package.

We cut stars from the frame using sigma clipping and manually cut a region with the nebula. The values in pixels were approximated using {\it scipy}'s \citep{2020SciPy-NMeth} 3-rd order smooth bivariate spline. The background subtracted region with nebula is presented in the middle panel of Figure~\ref{fig:morphology_obs}.

In order to perform quantitative comparison, we built profiles along the tail of the nebula (see Section~\ref{sec:profiles} for details) for both models and observational data. In case of observations, we worked with regions of previously cut nebula without stars, missing values were substituted with averages along column of pixels. In observational data, the waves have length 30'', which corresponds to 0.3~pc, for distance of 2~kpc (ATNF v1.70, \cite{2005AJ....129.1993M}). We linearly stretched our model to fit the observed nebula.
So for convenience, we present overlaid profiles in celestial coordinates for PSR J0742-2822 and in spacial ones for models with location of the second ring (first bubble) approximately matched. Signal-to-noise ratio was low, so we binned the observed profile with resolution of 3''. Model profiles were smoothed with the same on-picture width uniform filter. We see a systematic error in the observed profile rising from the head of the nebula to its tail (from almost zero to $\sim 40 \% $) due to low useful signal and high background variations symmetric around the axis of the nebula (possible ionisation halo visible in \Ha?).

In general, we have good agreement between models and observation. Mismatching of the head positions (model one lags behind rapid increase of observational profile on $x<0$, see Figure~\ref{fig:morphology_obs}) may be due to limitations of the numerical scheme. One possibility is the ionisation state can't be rendered on the narrow region between the shock and the contact discontinuity with a head-on stream of material. The flow in 2D models is more stable and ring structures are more pronounced compared to 3D case. 3D model should be more smooth. It will be addressed in future studies. 

The other possibility is absence of accounting for electron-ion equilibration processes on shock waves in {\it PLUTO}, where all particles are considered to be in equilibrium after passing the shockwave. \cite{2007ApJ...654L..69G} estimate that inequality of electron ($T_e$) to proton temperature ($T_p$) takes place for shock velocities more than $\approx 400$~km/s with $T_e / T_p \propto M^{-2}$, where $M$ is Mach number of a shock. In v03g43 model bubbles expand with velocities $\sim 150$~km/s, for which equilibration is rapid. The head of the nebulae moves through ISM with $V_{NS} = 450$~km/s rendering $T_e$ slightly less than $T_p$. For v1g43 and v1g53 models this effect can be stronger and result in some lines excited predominantly by protons and other nuclei at the head of the nebula. The similar scenario is observed at a fast expanding shell of SN~1006 remnant \citep{1996ApJ...472..267L}.

\section{Predictions of observational possibilities and features}\label{sec:predictions}
\subsection{Emissivity profiles along tail of nebula as quantitative description of its morphology}\label{sec:profiles}

\begin{figure*}
        \includegraphics[width=.49\textwidth, angle=-0, trim=0 0 0 0]{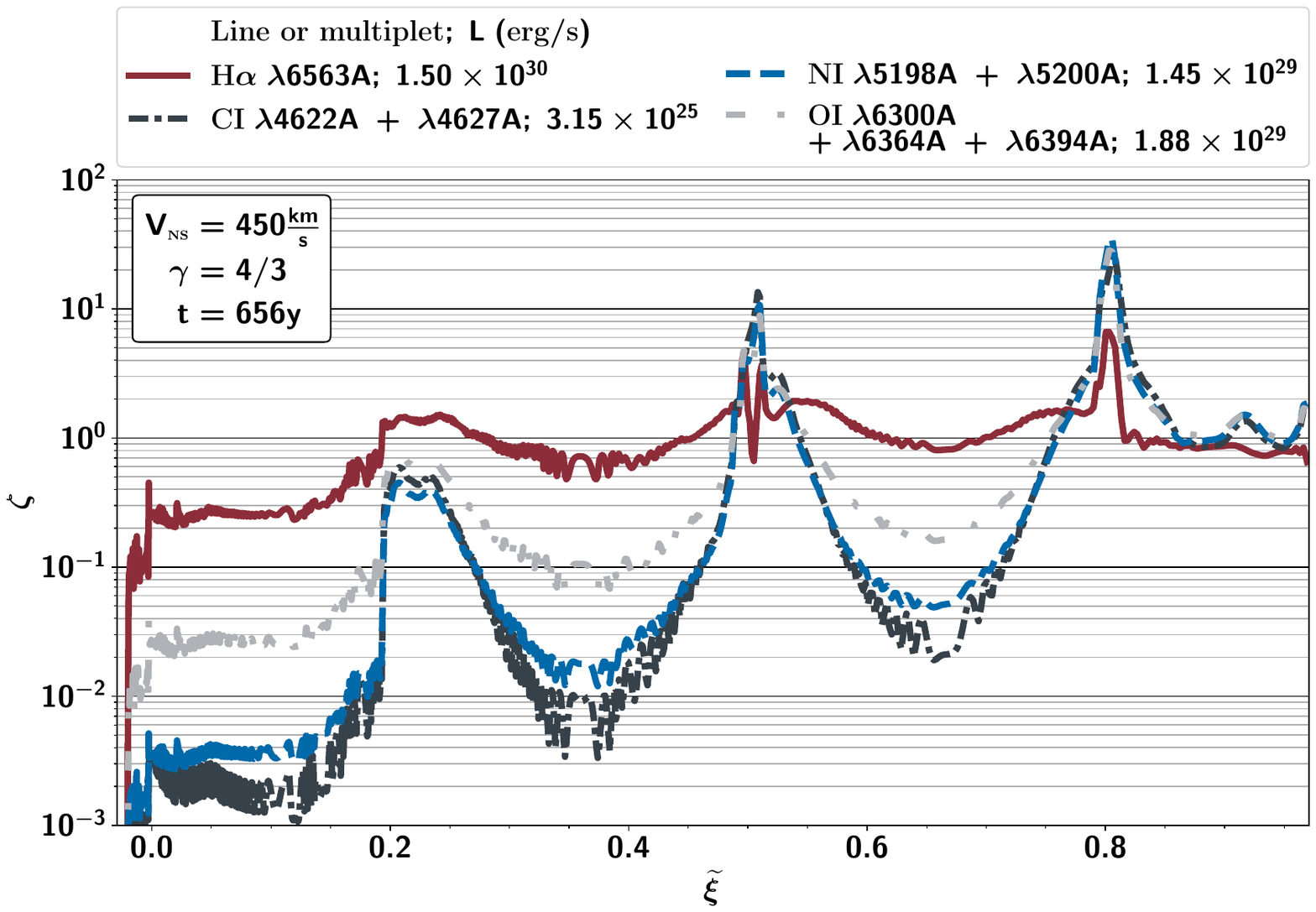}
        \includegraphics[width=.49\textwidth, angle=-0, trim=0 0 0 0]{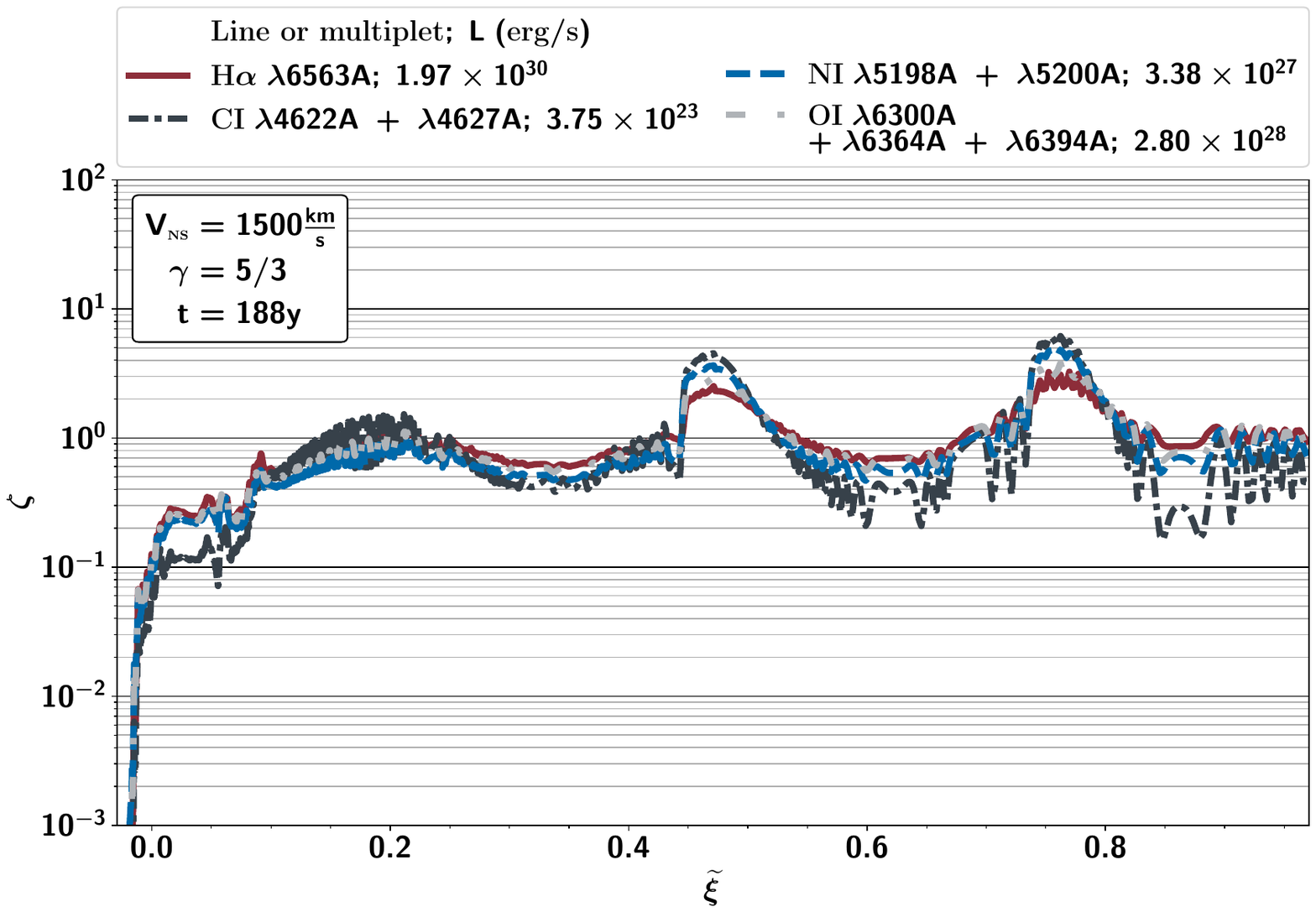}
        \includegraphics[width=.49\textwidth, angle=-0, trim=0 0 0 0]{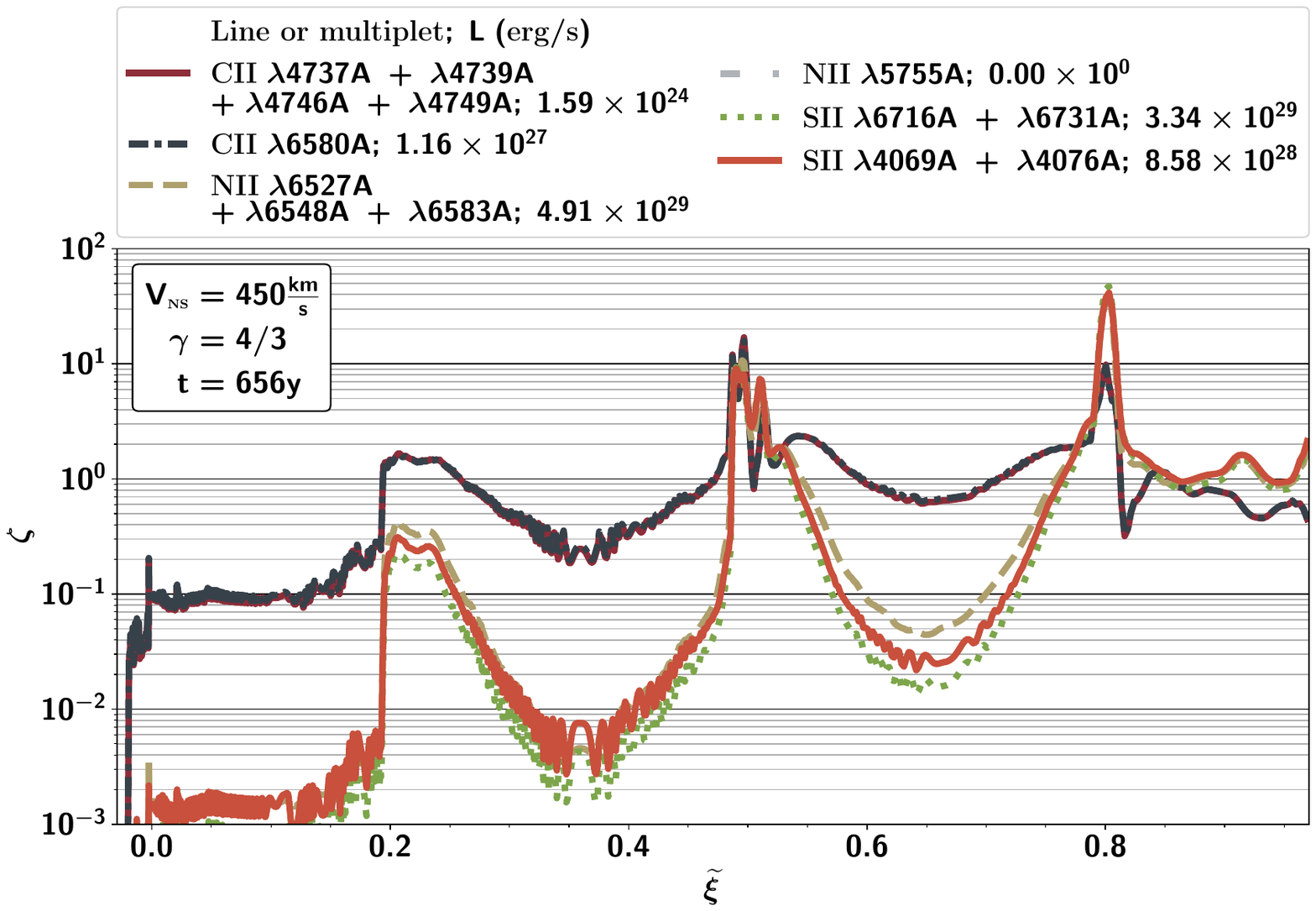}
        \includegraphics[width=.49\textwidth, angle=-0, trim=0 0 0 0]{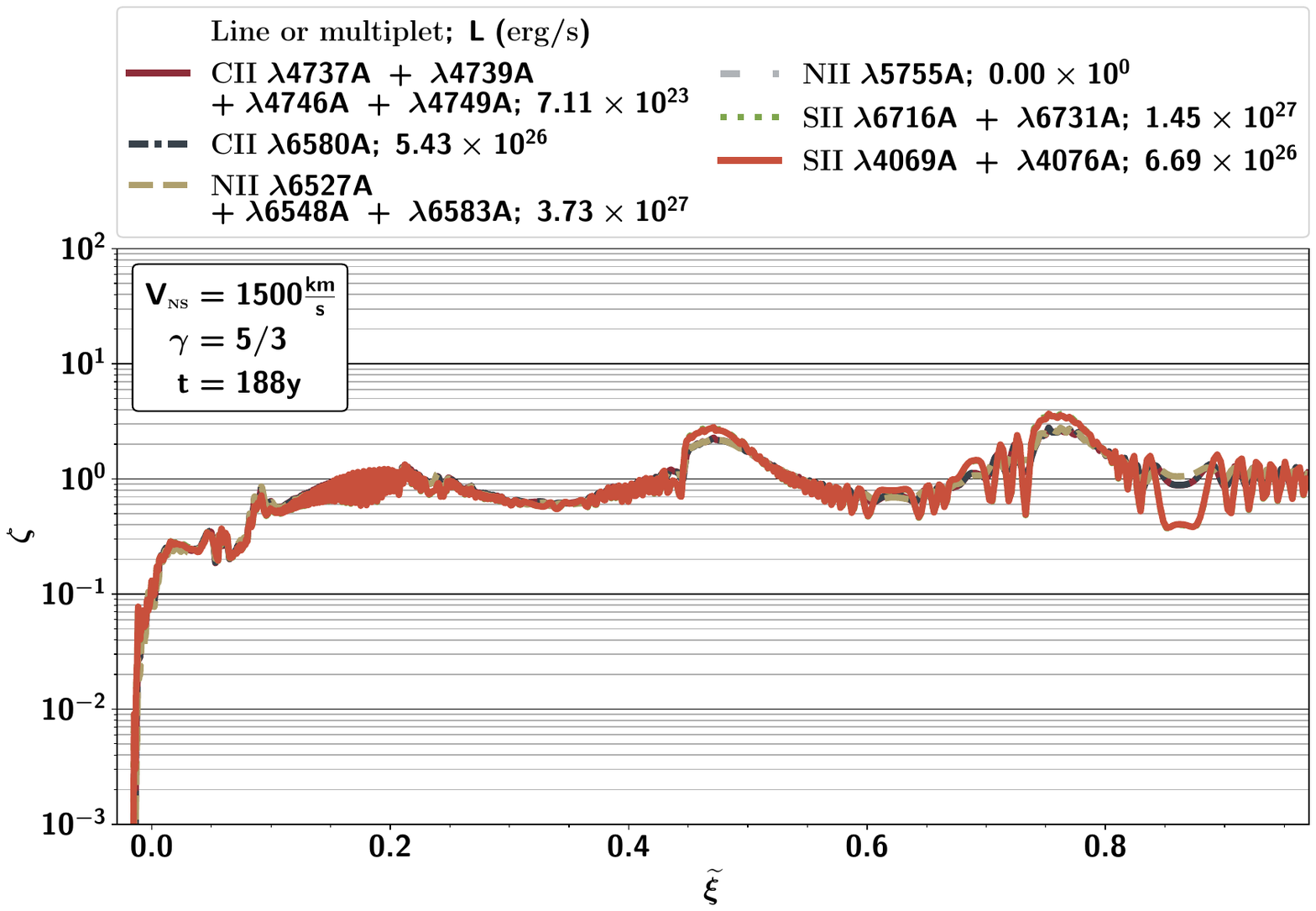}
        \includegraphics[width=.49\textwidth, angle=-0, trim=0 0 0 0]{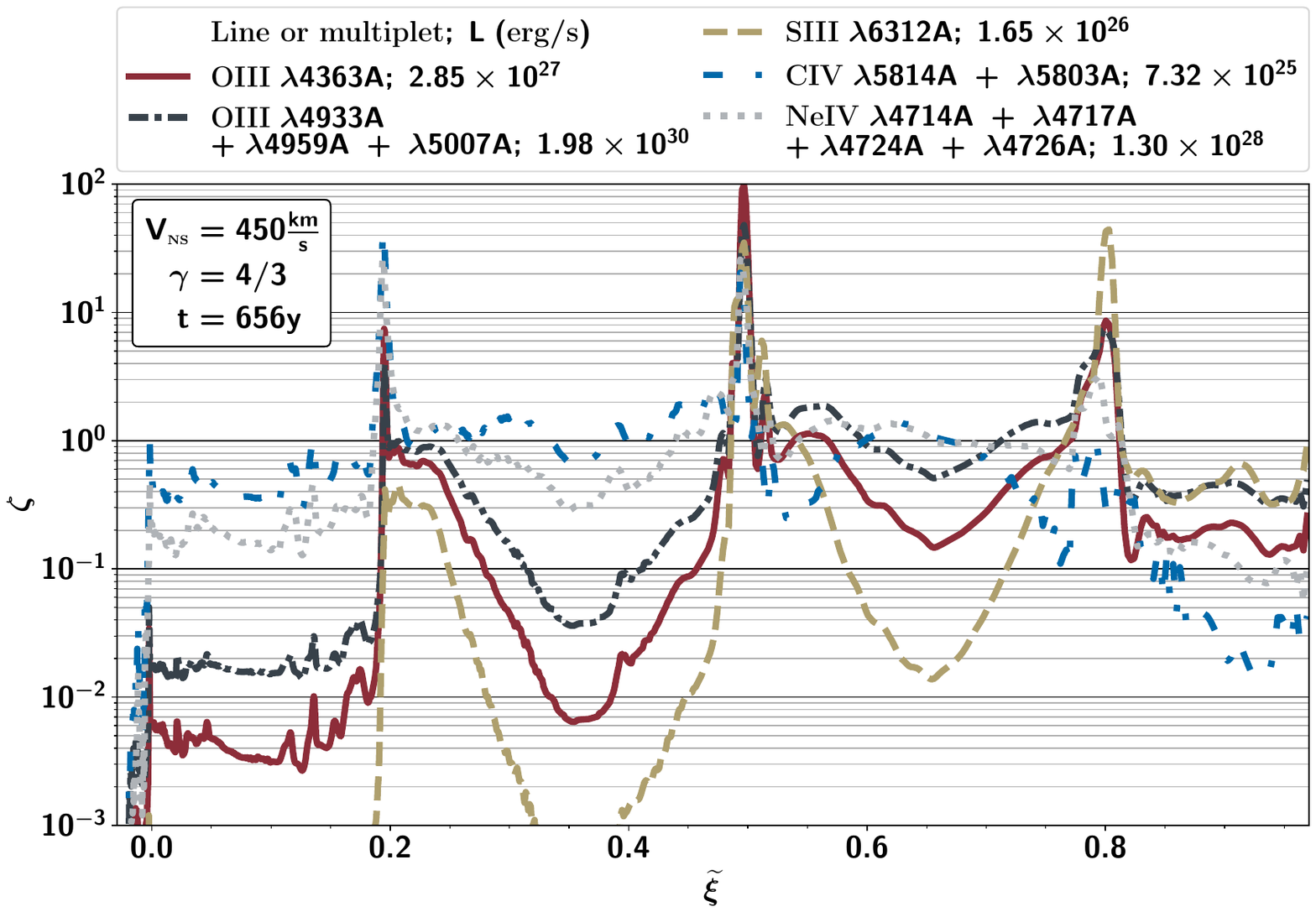}
        \includegraphics[width=.49\textwidth, angle=-0, trim=0 0 0 0]{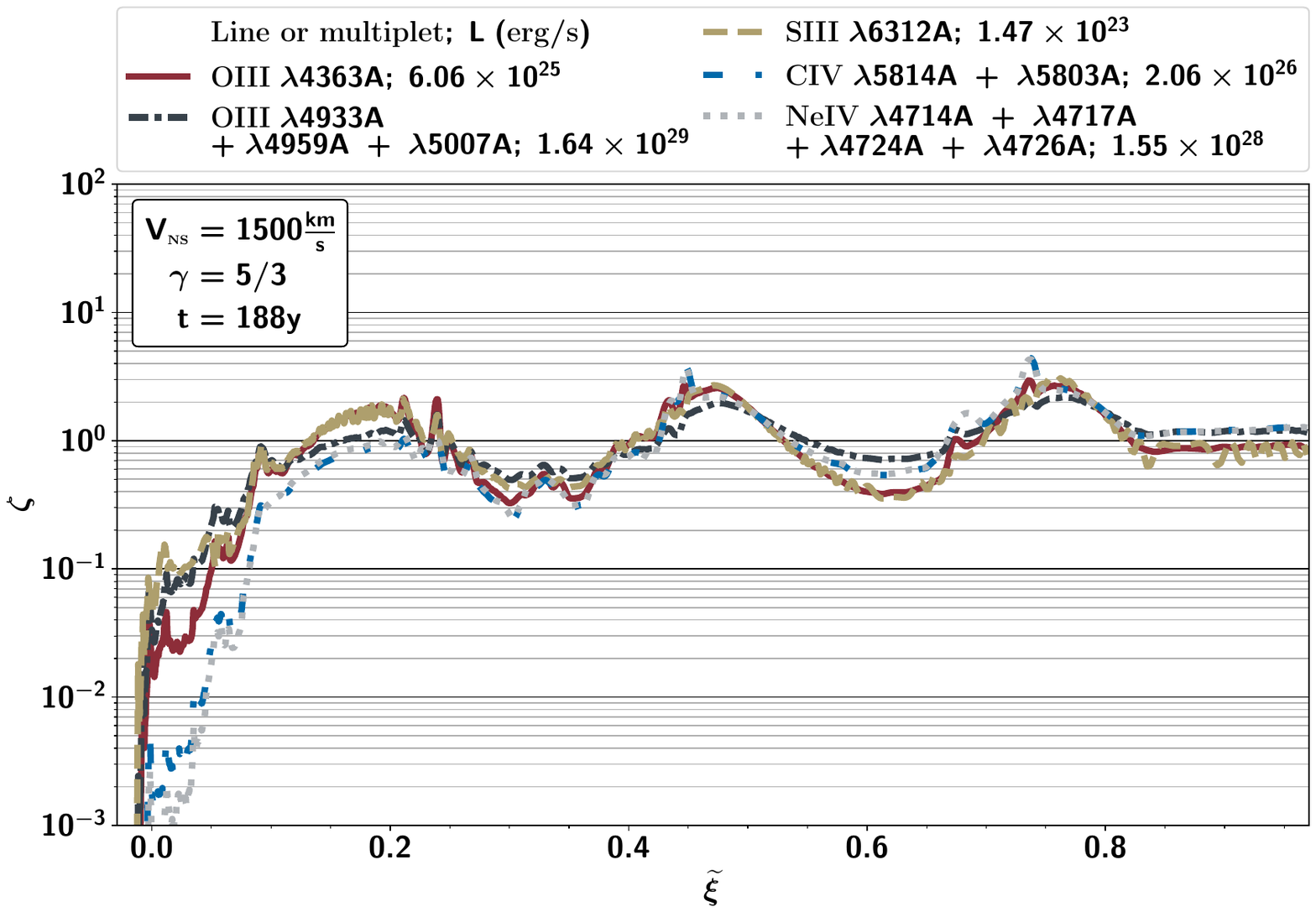}
        \caption{Profiles of model nebulae luminosity along tail for models v03g43 and v1g53 (the brightest and the dimmest rings, respectively). Lines are divided into groups by ionisation stages and shown on different plots. The luminosity of the nebula in lines is shown next to each of them in brackets.}        
        \label{fig:linesProfiles}
\end{figure*}

In order to analyse the contribution of different regions of the of nebula to its luminosity and compare models to each other, we built profiles along $z$-coordinate (see Figure~\ref{fig:linesLuminosity}). Coordinates and profiles were converted to dimensionless ones according to following expressions:
\begin{align}
        \xi                   & = \frac{z-z_{NS}}{z_{max}-z_{min}}~\text{-- dimensionless coordinate,}                                                                                                                          \\
        \begin{split}
                \zeta ({\tilde{\xi}}) & = \frac{\partial_{\tilde{\xi}}L(\xi<\tilde{\xi})}{L_{\text{full}}} =
                \\
                & = \frac{(z_{max}-z_{min})\int_{0}^{R_{max}}\eta_{em}(R,\tilde{z})Rdr}{\int_{z_{min}}^{z_{max}}dz \int_{0}^{R_{max}}\eta_{em}(R,z)Rdr},
        \end{split}
\end{align}
where $\tilde{\xi}=\xi(\tilde{z})$, $z_{NS}~(=0)$ -- pulsar location. $\zeta ({\tilde{\xi}})$ shows to what part of full luminosity in line the unit coordinate $\tilde{\xi}$ corresponds. For uniform distribution of luminosity along a nebula's tail, the equation holds: $\zeta ({\tilde{\xi}}) = 1$.

The profiles for \Ha\ are the closest to uniform luminosity distribution. In the same time, profiles of lines of triply ionised atoms (\CIVp, \NeIVp; see Figure~\ref{fig:linesProfiles}) show a spatial lag in growth comparing to others for high pulsar velocities. In v1g53 model, most of the considered lines reach a level of $\zeta = 0.1$ in  $\tilde{\xi} = 0$, while for \CIVdoublet\ doublet and \NeIVqaudruplet\ quadruplet that value amounts to $\tilde{\xi} = 0.1$. This difference corresponds to 10~a. So ISM starts emitting only in the tail. Such a lag is likely to be caused by the necessity of several ionisations of the atom before emission. In this case, the head of a nebula is expected to have the shape of an open tube.

There are coinciding peaks in all luminosity profiles, maximum values of which depend on the lines and the models. They correspond to rings in intensity maps. In \Ha\ these peaks are less noticeable, while in \OIIIp\ lines, for example, they are the most luminous parts of the nebula. The brightness of the rings is the lowest in the v1g53 model and the highest in the v03g43 model. Thus, presence of bright and high-contrast rings in the nebula in regions where the density of ISM is highest, may indicate pulsar having an intermediate value of velocity.

We expect a disappearance of some elements of the nebula morphology in some cases. In v03g43 and v03g53 models, the first bubble (head of nebula) has little contribution to overall luminosity of the nebula. This effect is stronger for lines of singly ionised atoms than for neutral ones: in v03g43 model $\zeta \in [1; 2] \times 10^{-3}$ for \NIIp\ and \SIIp\ lines, whereas $\zeta \in [2; 4] \times 10^{-3}$ for \CIp\ and \NIp\ lines, and even as high as $3 \times 10^{-2}$ in \OIp\ lines on the first bubble. The same feature is noticeable for the second bubble -- profiles of \NIIp\ and \SIIp\ lines lie under $2 \times 10^{-2}$ and \CIp\ and \NIp\ are above, \OIp\ reaches $0.1$. A possible reason is that bubbles, which are in the process of expansion, are smaller than fully developed ones (also a layer between bow shock and contact discontinuity is thinner), and have lower overall luminosity. In v01g43nv model, a similar effect is present -- in the same lines, the foremost part of nebula head is dimmer than other regions.

\subsection{Dependence of a nebula luminosity in various lines on pulsar velocity}
\label{sec:LinesLuminosity}

\begin{figure}
        \includegraphics[height=.3\textheight, angle=-0, trim=0 0 0 0]{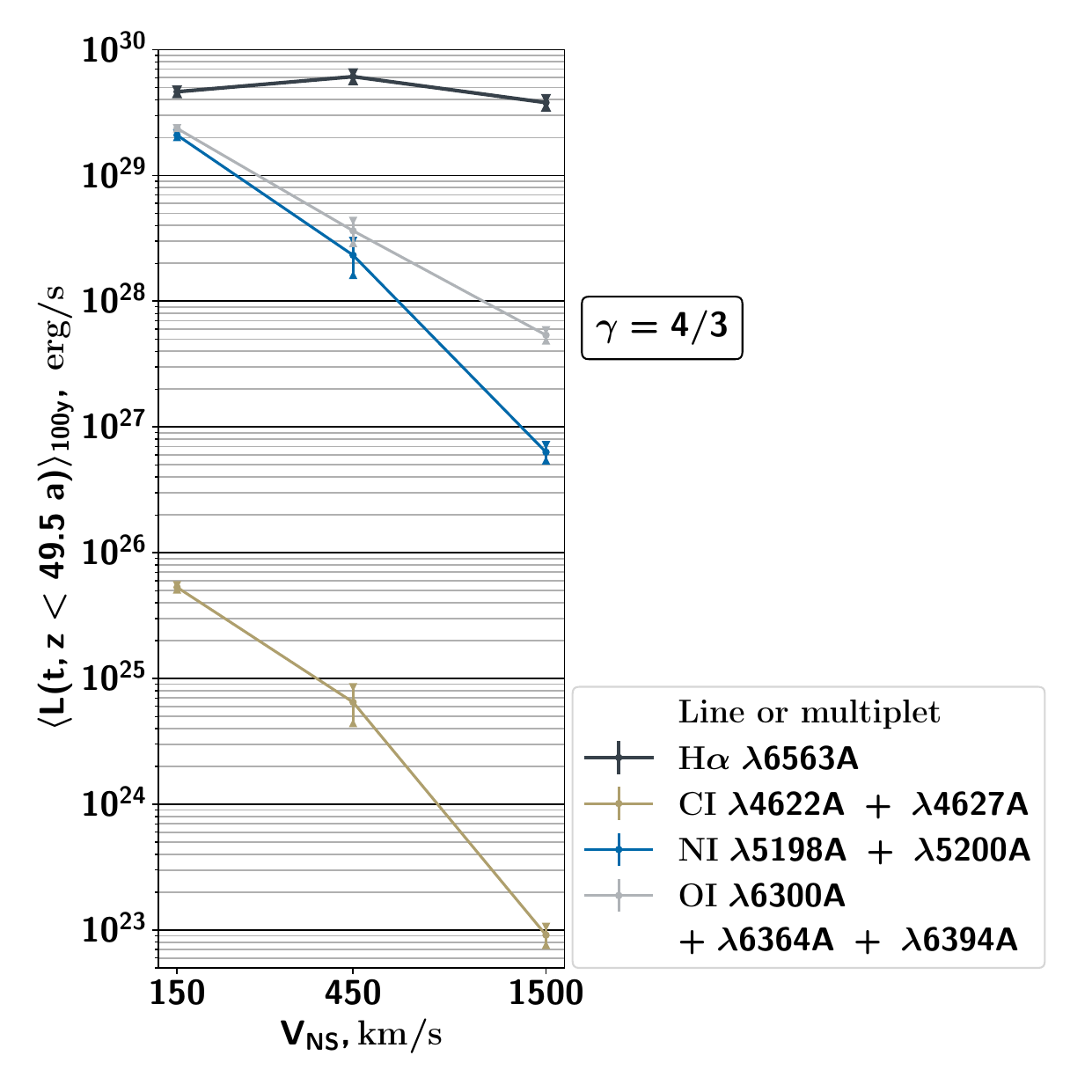}
        \includegraphics[height=.3\textheight, angle=-0, trim=0 0 0 0]{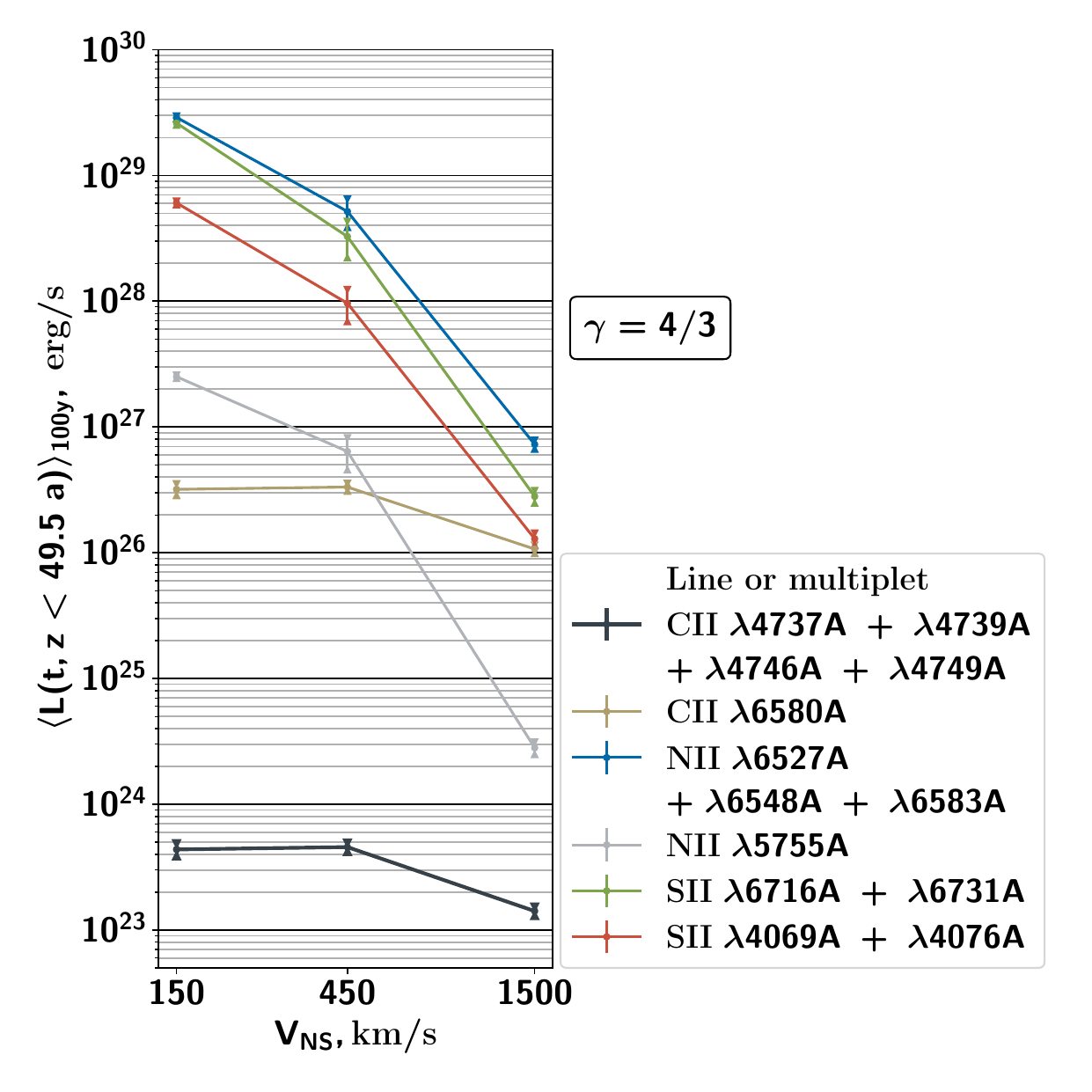}
        \includegraphics[height=.3\textheight, angle=-0, trim=0 0 0 0]{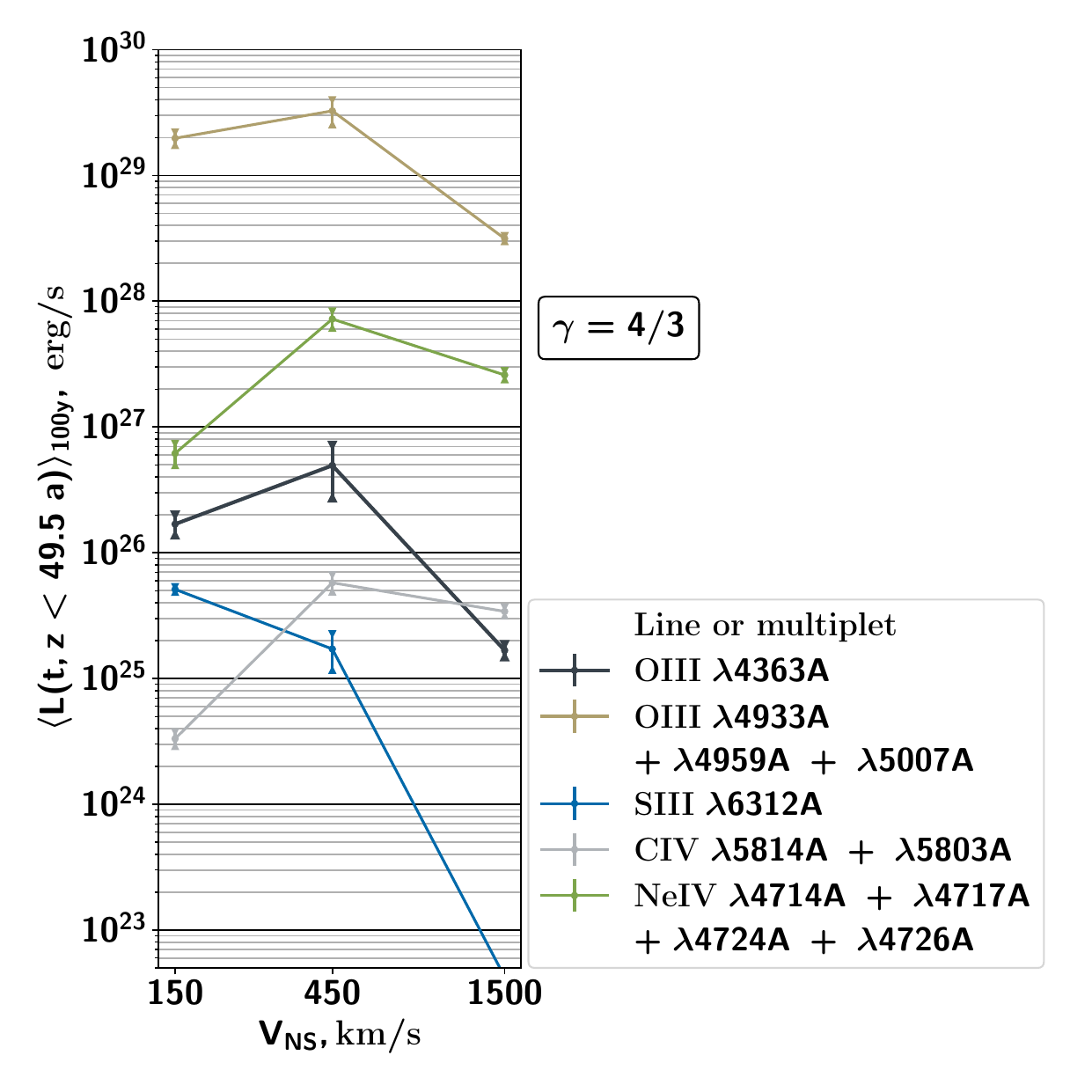}

        \caption{Model luminosity of the front part of the nebula ($z < 49.5$~a) averaged over last 100 years of the simulation versus pulsar velocity. We presented cases of ultrarelativistic gas ($\gamma = 4/3$, models v01g43nv, v03g43, v1g43). Lines are divided into groups by ionisation stages and shown on different three panels. Error bars show RMS of luminosity.}
        \label{fig:linesLuminosity}
\end{figure}

We plotted the averaged luminosity of nebulae depending on pulsar velocity (see Figure~\ref{fig:linesLuminosity}). Because in v01g43nv model morphology developed up to $z \approx 50$~a, we limited this analysis to $z = 49.5$~a. Luminosity was averaged during the last 100 years of the simulation, error bars correspond to its root-mean-square (RMS) during a given time. We compared models with different adiabatic index (both $\gamma = 4/3$ and $5/3$) on full length of nebula ($z \le 100$~a) in \ref{sec:LuminosityBothGamma}. We show light curves of the model nebulae in \ref{sec:LightCurves} to illustrate luminosity variability in models. The nebula luminosity strongly depends on pulsar velocity. It can be explained by the ionisation stage variation of the element which emits lines. It allows us to understand at which conditions bow-shock PWNe can be bright.

Most lines of neutral atoms (upper panel of Figure~\ref{fig:linesLuminosity}) show the strongest monotonic and almost power law dependency. \OItriplet\ triplet is more than 1.5 order brighter for $V_{NS}=150$~km/s than for $1500$~km/s. This line is also the brightest multiplet of neutral atoms at $2 \times 10^{29}$~ergs/s. A second bright multiplet of almost the same luminosity is \NIdoublet, the difference in brightness here is even more and equals 2.5 orders of magnitude. \CIdoublet\ doublet is much more faint at $5 \times 10^{25}$~ergs/s and shows even higher difference in brightness at almost 3 orders of magnitude. We can claim that \OIp\ and \NIp\ lines are favourable for observations with luminosity very close to \Ha, but potential target nebulae must contain only low-velocity pulsars (around $150\text{km/s}$). The exception here is \Ha\ line, which luminosity is almost constant and varies in range of $[4;~7] \times 10^{29}$~ergs/s. High luminosity in  \Ha\  is natural, because of hydrogen is the most abundant element. 

Lines of singly ionised atoms (middle panel on Figure~\ref{fig:linesLuminosity}) show the same trend, with up to almost 3 orders of magnitude differences in luminosity between velocities. The dependency doesn't resemble power law like in case of neutral atoms and the ``knee'' begin to emerge in plots at $V_{NS}=450$~km/s. The dependency is monotonic, still with steeper slope at higher pulsar velocity. These lines are favourable for observations with low pulsar velocity also. The brightest sets of lines are  \NIItriplet\ triplet and \SIIdoubletMain\ doublet, both peaking at around $3 \times 10^{29}$~ergs/s. In the most lines, ``knee'' is present at almost the same degree, except for faint ($L < 4 \times 10^{27}$~ergs/s) \CIIp\ lines.
There it transforms to ``plateau'' with constant luminosity between $V_{NS}=150$ and $450~\text{km/s}$. Decrease after $V_{NS}=450\text{km/s}$ is also lower than of other lines.

Luminosity of doubly and triply ionised atoms lines vs velocity is no longer monotonic with maximum at $V_{NS}=450\text{km/s}$. The tendency is following, the higher ionisation stage the higher velocity required for the peak. The \OIIItriplet\ nebular triplet of  is brightest and is in perspective for detection. Luminosity in these lines are about $2 \times 10^{29}$~ergs/s (peaking at $3 \times 10^{29}$~ergs/s) from $V_{NS}=150$~km/s till $550$~km/s, which makes nearly all known bow-shock PWNe potential candidates for observations. Together with concentration of luminosity in rings, this makes \OIIIp\ lines quite promising. 

\section{Discussion and Conclusion} \label{sec:conclusion}

In this work, we combine hydrodynamic simulations and non-LTE modelling of line transitions of atomic and ionic species to produce expected intensity maps that allow to reconstruct both density and chemical composition structure in ISM at ultra-small scales. 
         
We developed {\it Shu} \citep{Shu2023} program package. It allows calculation of non-LTE intensity maps (analogs of frames in narrowband filters) in more than 150 spectral lines of H, He, C, N, O, S, Ne atoms and their ions for different angles between model axes and picture plane. We used {\it Shu} to build synthetic intensity maps of model bow-shock PWNe in various optical lines (listed in Table~\ref{tab:ListOfLines}).

A particularly promising application of the present work is to the interaction of the fast-moving pulsar with dilute warm component of ISM, which otherwise is hard to observe. We demonstrate that in this case, one expects relatively bright line emission.

As neutral hydrogen propagates through a nonradiative shock, the \Ha\ emission line exhibits linear polarisation. This effect stems from anisotropic excitation by fast-moving electrons and protons, a mechanism supported by observations of SN 1006 \citep[see ][]{2015ApJ...815L...9S}. In the case of PWN, the characteristic shock speed is about 100~km/s, so the polarisation degree should be less than 0.1\% \citep[see][]{1990ApJ...362..219L}.

For the majority of observed objects, there are only two significant free parameters: external density structure and angle between pulsar velocity and picture plane. The second one can potentially be found from panoramic spectroscopy \citep[see][]{2022ApJ...939...70D}  or from direct morphology comparison with models, while the first one is a major question of interest. As we provide the first direct comparison between results of hydrodynamic modelling and observations of \cite{2014ApJ...784..154B}, the road becomes open to tune the model parameters to reconstruct the structure of ISM inhomogeneities and clouds. 

Various lines are formed by ISM inhomogeneities in different locations, showing various morphological features. This fact can be used to reconstruct the abundance structure of different elements. There is also a possibility to obtain the abundance structure of elements using data from different ions (for example, O from \OIp\ and \OIIIp, N from \NIp\ and \NIIp). Realisation of this possibility can be done after successful detection of bow-shock PWNe in several lines and can be a unique source of information about warm component of ISM.

We predict the expected features of bow-shock PWNe morphology in various spectral lines and built profiles along the tail for their quantitative description. Peaks in all luminosity profiles correspond to rings on intensity maps. In \Ha\ these peaks are less noticeable, while in \OIIIp\ lines they contain the most part of the nebula luminosity. Profiles of lines of triply ionised atoms (\CIVp\ and \NeIVp) show a spatial lag in growth comparing to others for high pulsar velocities. In this case, a head of a nebula is expected to have a shape of an open tube. Seeming disappearance of some elements of nebula morphology due to low intensity relative to other regions is expected in some cases.

We predict expected luminosity and favourable conditions for observations in spectral lines of optical range. \OItriplet, \NIdoublet, \NIItriplet\ and \SIIdoubletMain\ are expected to be bright in nebulae of relatively slow pulsars (up to $3 \times 10^{29}$~ergs/s for $V_{NS}=150$~km/s), but aren't expected to be observable at higher velocities. \OIIItriplet\ lines are expected to be bright ($[2;~3]\times 10^{29}$~ergs/s) from $V_{NS}=150$~km/s to $550$~km/s, what makes them the most promising candidate for observations. This makes bow-shock PWNe are potential targets for observations by both earth-based 4m+ telescopes and space telescopes.

We calculated five 2D relativistic hydrodynamic models of bow-shock PWNe with detailed accounting for ionisation state of H, He, C, N, O, S and Ne. We considered pulsar velocities of $V_{NS}=150$~km/s, $450$~km/s, and $1500$~km/s and adiabatic index in ultrarelativistic ($\gamma=4/3$) and classical ($\gamma=5/3$) limit. Periodic variation of interstellar gas density due to inhomogeneities of ISM were included. 

We compared our models to existing observations of bow-shock PWNe. We scaled optical fluxes in \Ha\ \citep{2014ApJ...784..154B} with ANTF \citep{2005AJ....129.1993M} data from radio spectral range. Scaled luminosity in most cases show coincidence with models with accuracy of about 30\%. Dependency of nebula luminosity from pulsar velocity obtained from models puts some constraints on flux scaling laws ($\epsilon_{\text{H}\alpha} \eta^2 \propto V_{NS}^{-1}$). We compared morphology of model nebulae with nebula of PSR J0742-2822. Features of model profiles along nebula tail and observed ones are very similar. Profile of v03g43 model makes the best fit with accuracy of about 30\% on most areas.

Despite on overall good agreement of modelled nebulae and observed ones in integrated H$_{\alpha}$ flux, there are inconsistencies for several objects, such as J1856-3754, PSR J2225+6535 and in some degree for PSR J0437-4715. However, this fact can be used as a marker of peculiarities of these objects. For example, J1856-3754 is a member of ``great seven'' and, probably, forms thermal radiation ionisation dominated nebulae \citep{2015SSRv..191..171P}.

\begin{acknowledgement}
The authors appreciated to the anonymous referee for the constructive comments. The simulations were performed on CFCA XC50 cluster of National Astronomical Observatory of Japan (NAOJ) and RIKEN HOKUSAI Bigwaterfall. We thank Alexey Moiseev and Alexander Kolbin for useful discussion and valuable suggestions, Roger Romani for kindly providing the use of observational data of PSR J0742-2822. We acknowledge using python packages numpy \citep{harris2020array}, mpi4py \citep{9439927}, pyCUDA \citep{kloeckner_pycuda_2012}, astropy \citep{2022ApJ...935..167A}, reproject\footnote{URL: https://reproject.readthedocs.io}, scipy \citep{2020SciPy-NMeth}, matplotlib \citep{Hunter:2007}, tueplots\footnote{URL: https://github.com/pnkraemer/tueplots} and SymPy \citep{10.7717/peerj-cs.103}. We used Paraview software \citep{Paraview} to plot density maps and streamlines.
\end{acknowledgement}

\paragraph{Funding Statement}

This research was supported by the grant 23-22-00385 of the Russian Science Foundation. I.N.~Nikonorov acknowledges partial support from Gennady Komissarov Foundation (Appendixes). 

\paragraph{Competing Interests}

None.

\paragraph{Data Availability Statement}

The data underlying this article will be shared on reasonable request to the corresponding author.

\printendnotes

\printbibliography

\newpage

\appendix

\section{Calibration methods}
\label{sec:cal}

Analytical dependencies of \Ha\ bow-shock PWNe luminosity from various parameters were investigated in  \citep{1993Natur.362..133C, 2002ApJ...575..407C, 2014ApJ...784..154B}. Various physical assumptions caused complex scaling laws from pulsar velocity. In contrary, we aimed to provide the most general scaling as possible.

Consider an average amount of \Ha\ quanta radiated by one neutral hydrogen atom after passing through a strong shock wave ($\epsilon_{\text{H}\alpha}$). \cite{1991PASP..103..781R} estimated the rule of thumb as:
\begin{equation}\label{eq:epsilonRaymond}
        \epsilon_{\text{H}_\alpha} \approx 0.2.
\end{equation}

\cite{2014ApJ...784..154B} analyse results of numerical simulations of \citep{2007ApJ...654..923H}, which is estimating radiation of \Ha\ quanta on shock waves. The authors found higher yield for $V_{NS} < 10^3$~km/s in assumption of electron-ion equilibrium behind the shock wave:
\begin{equation} \label{eq:epsilon_romani1}
        \epsilon_{\text{H}_\alpha}(V_{NS}) \approx 0.6 V_{NS,~7}^{-1/2}.
\end{equation}
The authors point out lower yield in non-equilibrium case, which take place at lower pulsar velocities:
\begin{equation} \label{eq:epsilon_romani2}
        \epsilon_{\text{H}_\alpha}(V_{NS}) \approx 0.04 V_{NS,~7}^{3/4}.
\end{equation}
Efficiency in Equations~(\ref{eq:epsilon_romani1}) and~(\ref{eq:epsilon_romani2}) equalises on $V_{NS} = 8.7 \times 10^2$~km/s and equals to $\epsilon_{\text{H}_\alpha} = 0.2$, the estimation of \cite{1991PASP..103..781R}. Following Equations~(\ref{eq:epsilon_romani1}) and~(\ref{eq:epsilon_romani2}), for models calculated in this work the efficiency should be 0.05, 0.12 and 0.15 for $V_{NS}= 150$~km/s, 450~km/s and 1500~km/s respectively.

We notice that for various conditions behind the shock wave $\epsilon_{\text{H}\alpha}$ estimate is either about constant, or some function of ISM material velocity in the shock wave's frame of reference (which is the same as $V_{NS}$ in this work). Serving the purpose of getting the most general law not assuming specific conditions on both sides of the shock wave, we have:
\begin{equation}
        \epsilon_{\text{H}_\alpha} = f_1 (V_{NS}),
\end{equation}
where $f_1 (V_{NS})$ -- some function of pulsar velocity.

\cite{2019MNRAS.484.4760B} give detailed description of bow-shock PWNe morphology. Consider stand-off distance, which characterise overall size of nebula. The classic equation connecting it with parameters of pulsar and ISM:\
\begin{equation}\label{eq:rs_short}
        r_s = \sqrt{\frac{L_w}{4 \pi c \rho_{ISM} V^2_{NS}}},
\end{equation}
where $\rho_{ISM}$ -- local density of ISM, $L_w$ -- pulsar luminosity or pulsar spin-down power ($\dot{E}$). 

Number of hydrogen neutral atoms passing the bow shock during unit time is proportional to ISM density, velocity of pulsar and the area of nebula's emitting layer ($S$), which in its turn proportional to stand-off distance squared ($r_s^2$) with some coefficient($\eta$):
\begin{align}
        L _{\text{H}_\alpha} &= \epsilon_{\text{H}\alpha} \rho_{ISM} V_{NS} S \\
        S &= \left(\eta r_s\right)^2.
\end{align}

In the last case, the proportionality coefficient ($\eta$) depends on the velocity of the pulsar. \cite{1993Natur.362..133C} propose linear dependency:
\begin{equation}\label{eq:propCordes}
        \eta \approx 30 V_{NS, 7}.
\end{equation}

\cite{2002ApJ...575..407C} suggest more general one with power law:
\begin{equation}\label{eq:propChattergee}
        \eta \propto V_{NS}^\beta,
\end{equation}
where $\beta$ is constant, with conclusion $\beta =  1$ is plausible. This and Equation~(\ref{eq:epsilonRaymond}) lead to $L_{\text{H}_\alpha} \propto \dot{E} V_{NS}$.

In order to compare these results with modelling, we consider even more general case with coefficient of proportionality $\eta$ being an arbitrary function of pulsar velocity:
\begin{equation}
        \eta = f_2(V_{NS}).
\end{equation}

In total, we have
\begin{align}
        \begin{split}
                L _{\text{H}_\alpha} &= \epsilon_{\text{H}\alpha} \rho_{ISM} V_{NS} S = \\
                &= f_1(V_{NS}) \rho_{ISM} V_{NS} r_s^2 f_2^2(V_{NS}) =\\
                &=\frac{L_w}{V_{NS}^2} V_{NS} f_1(V_{NS}) f_2^2(V_{NS});
        \end{split} \\
        L _{\text{H}_\alpha} & = \dot{E} f(V_{NS}), \label{L_VNS}
\end{align}
where $f(V_{NS}) = V_{NS}^{-1} f_1(V_{NS}) f_2^2(V_{NS})$ -- function of $V_{NS}$. It is unknown but is supposed to be common among observed nebulae, disregarding differences in chemical composition and contribution from individual features of morphology.

Thereby, we attained an equation for \Ha\ luminosity calibration to model one for given pulsar velocity:
\begin{equation}\label{eq:calibration}
        L_{\text{H}\alpha}^{calibrated} (V_{NS}) = L_{\text{H}\alpha} \dot{E}_{model}(V_{NS}) / \dot{E}.
\end{equation}
Here $\dot{E}_{model}$ -- spin-down power of model pulsar. This quantity corresponds to model $\rho_{ISM} = 1~m_p/\text{cm}^3$ and $r_s = 0.57$~a, and given $V_{NS}$, according to Equation~(\ref{eq:rs_short}).

\section{Luminosity comparison between models}\label{sec:LuminosityBothGamma}
\begin{figure}
        \includegraphics[height=.3\textheight, angle=-0, trim=0 0 0 0]{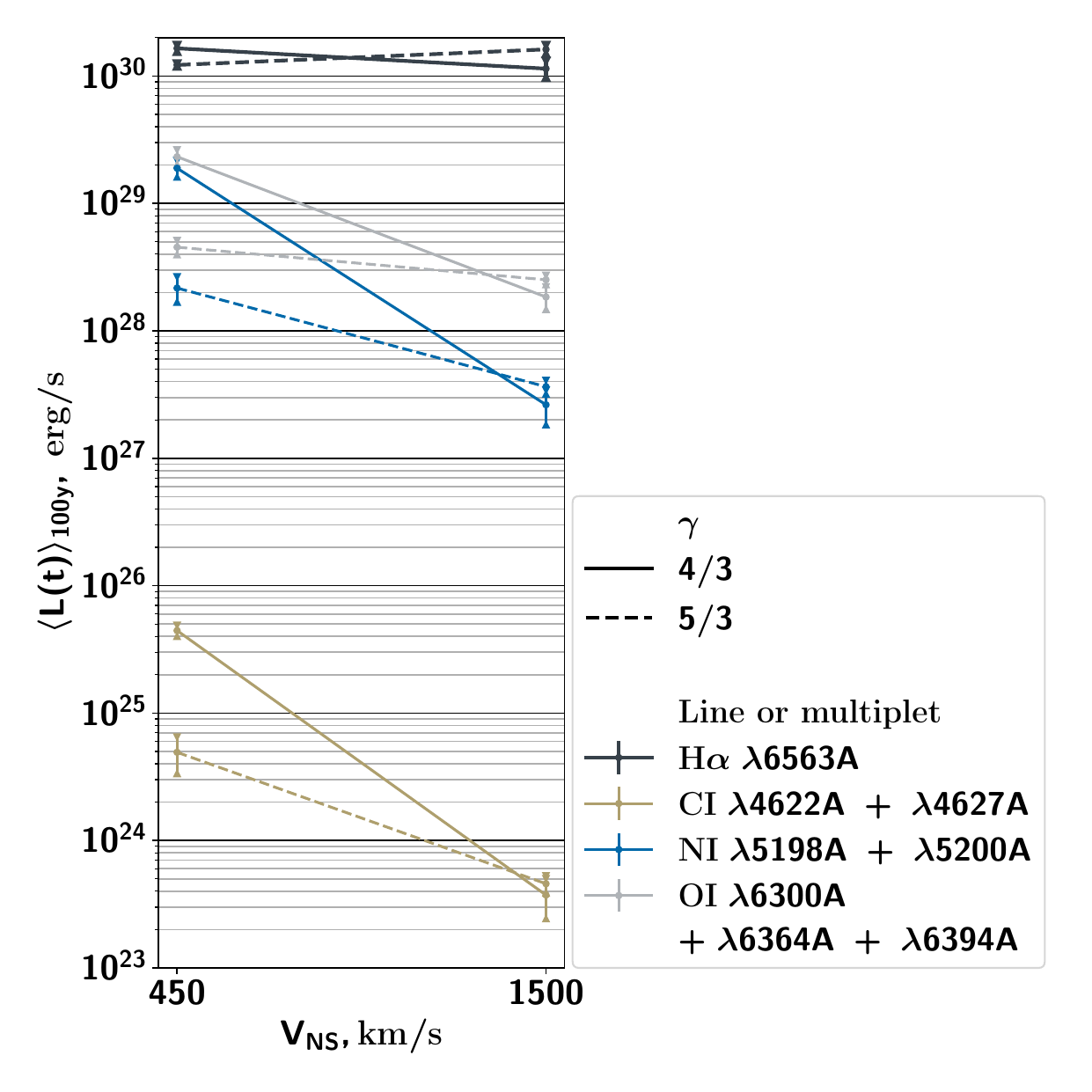}
        \includegraphics[height=.3\textheight, angle=-0, trim=0 0 0 0]{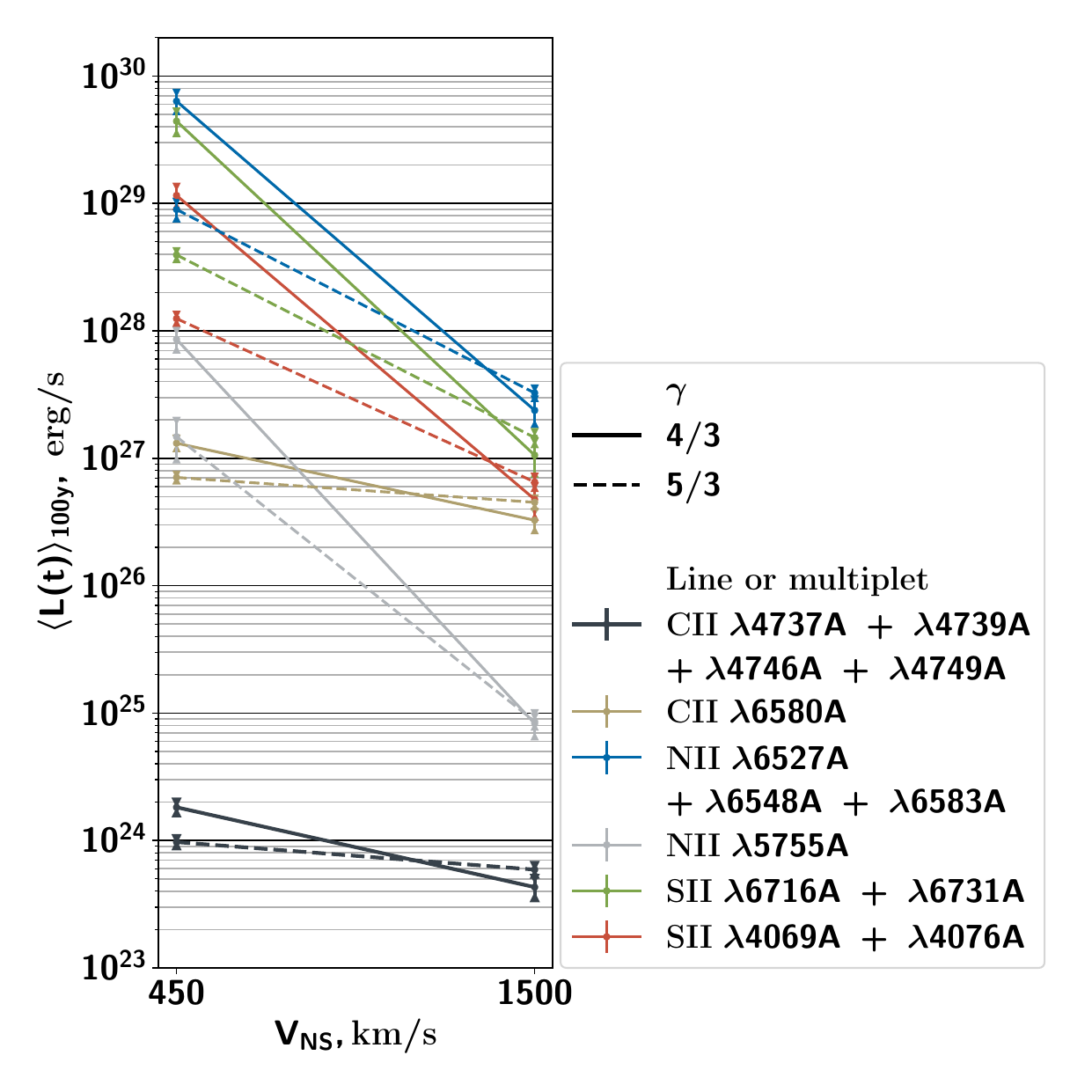}
        \includegraphics[height=.3\textheight, angle=-0, trim=0 0 0 0]{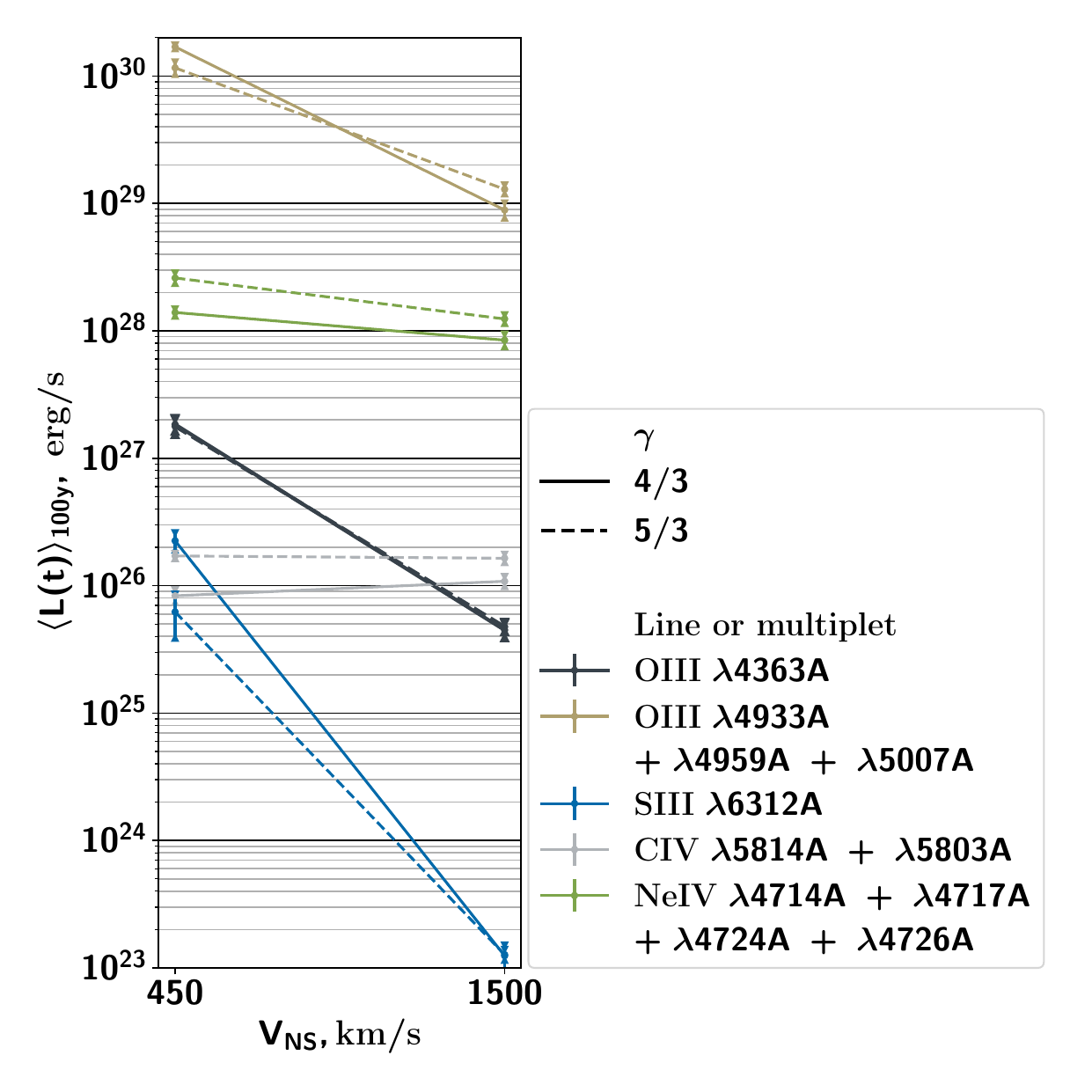}

        \caption{Model luminosity of the nebula ($z < 100$~a) averaged over last 100 years of simulation versus pulsar velocity. Here cases of intermediate and high velocities with adiabatic index in both of ultrarelativistic and classical limits are presented (models v03g43, v03g53, v1g43, v1g43). Lines are divided into groups by ionisation stages index and presented on different plots. Error bars show standard deviation of luminosity during time of averaging.}
        \label{fig:linesLuminosityBothGamma}
\end{figure}

On Figure~\ref{fig:linesLuminosityBothGamma} we present a luminosity comparison between models with adiabatic index in ultrarelativistic ($\gamma = 4/3$) and classical ($\gamma = 5/3$) limits on the full length of nebula tail ($z < 100$~a).

For high pulsar velocity ($V_{NS} = 1500$~km/s) the difference between luminosity in the same line between models with different adiabatic index is less than variations of luminosity. For intermediate value of velocity ($V_{NS} = 450$~km/s) the situation is the same with \Ha\ and \OIIIp\ lines. 

The difference is distinctly larger (up to an order of magnitude) in the cases of such bright lines as \NIp, \OIp, \NIIp\ and \SIIp. However, this difference is due to the morphology of the nebula. With $\gamma = 5/3$ rings are bigger and doesn't form well in model domain (see Figure~\ref{fig:rhoV1} and~\ref{fig:rhoV2}). This trait is caused by limitation of numerical scheme and consequently insufficient compression of the relativistic pulsar wind. In real nebulae, we expect rings to form and be visible, even with lower density, than in the case of $\gamma = 4/3$.

\section{Light curves of the model nebulae}\label{sec:LightCurves}

\begin{figure*}
        \includegraphics[width=.49\textwidth, angle=-0, trim=0 0 0 0]{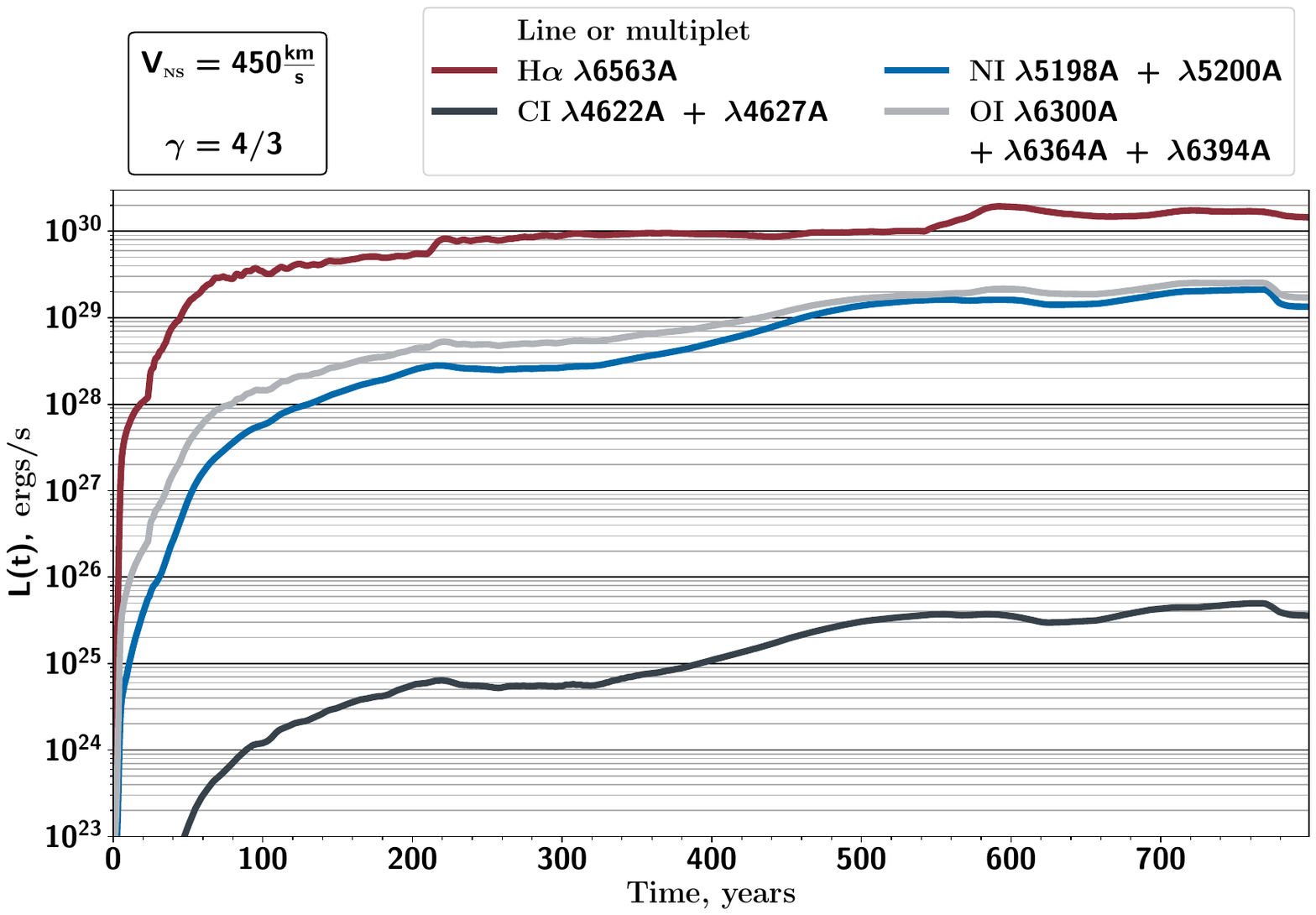}
        \includegraphics[width=.49\textwidth, angle=-0, trim=0 0 0 0]{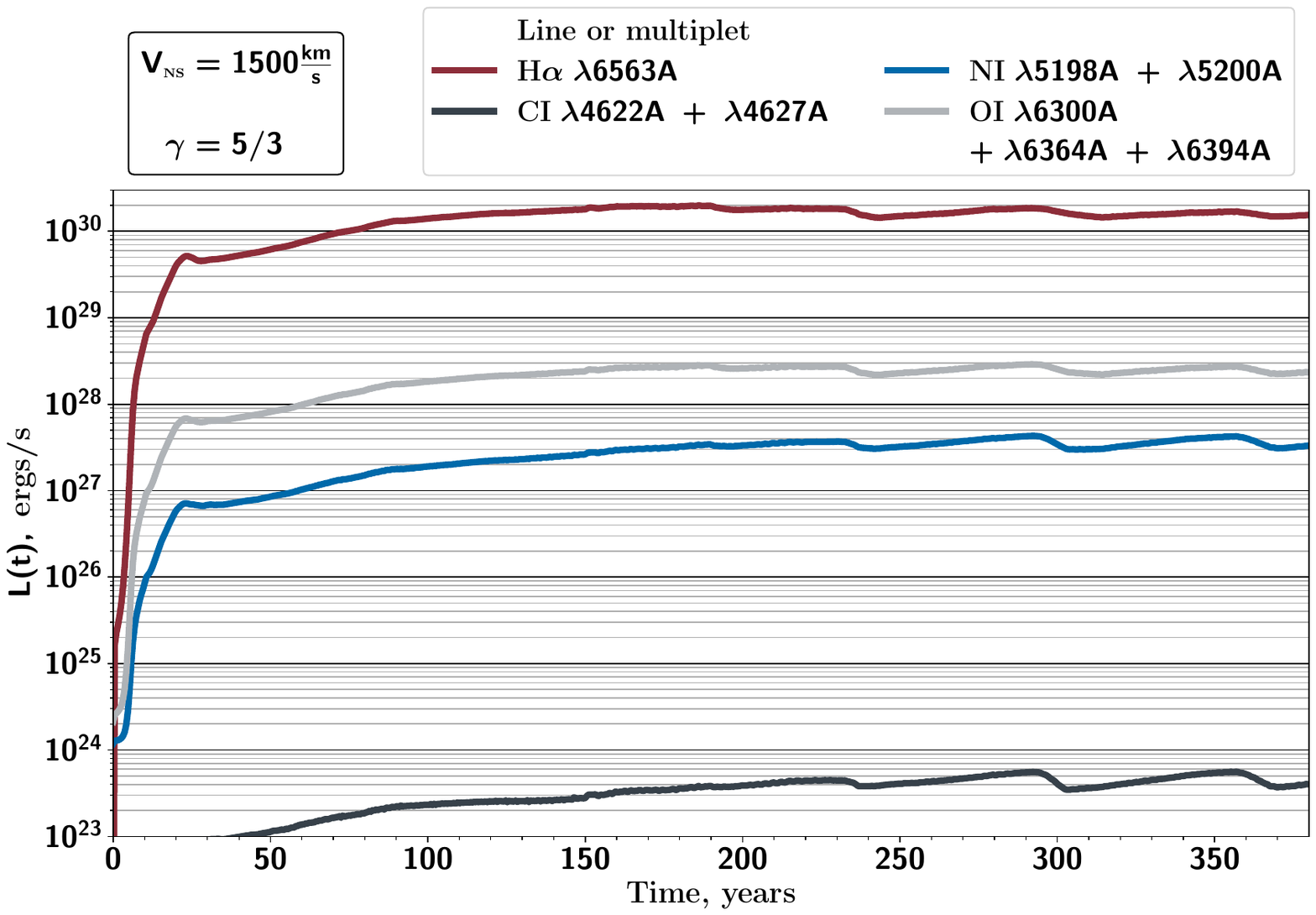}
        \includegraphics[width=.49\textwidth, angle=-0, trim=0 0 0 0]{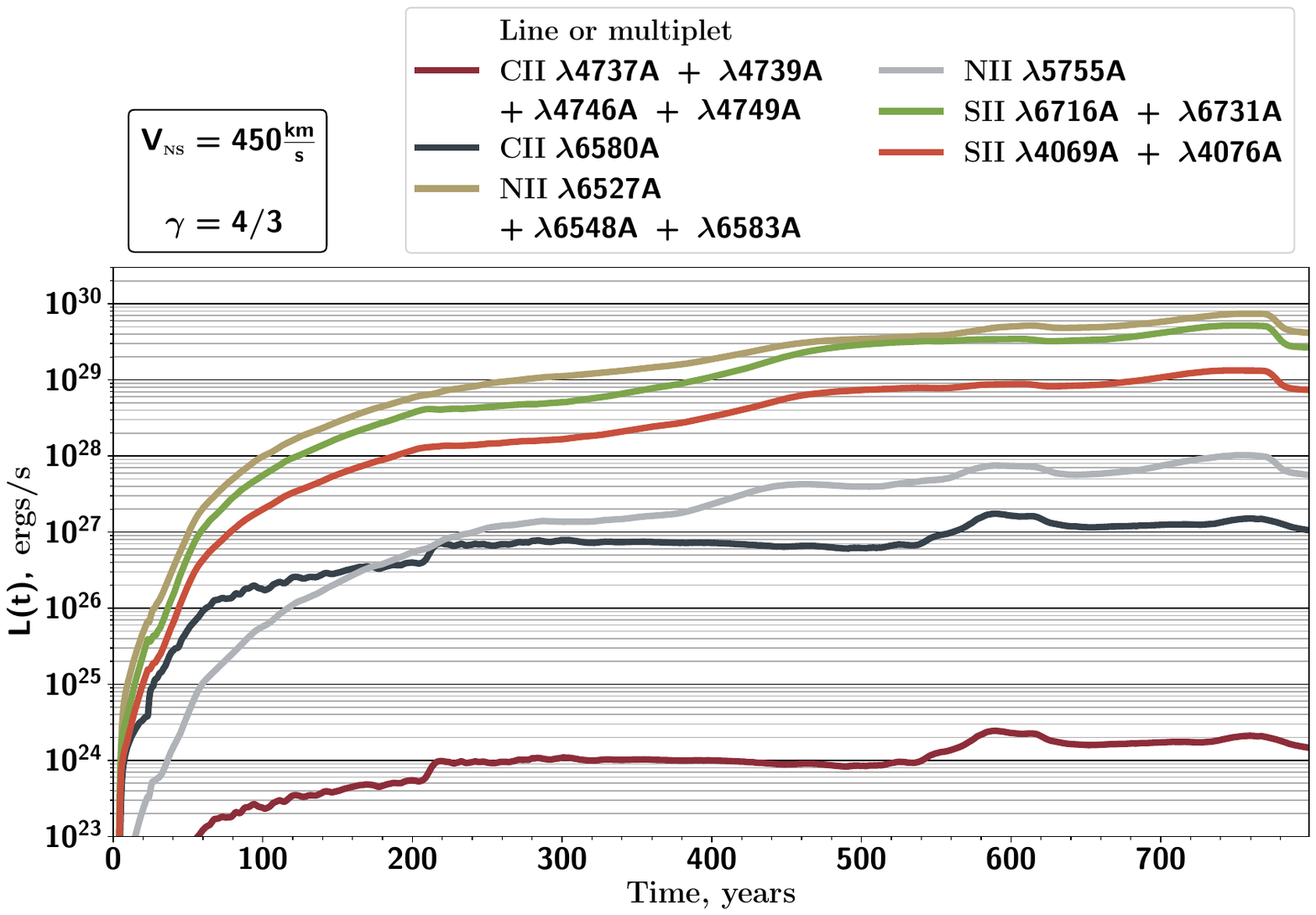}
        \includegraphics[width=.49\textwidth, angle=-0, trim=0 0 0 0]{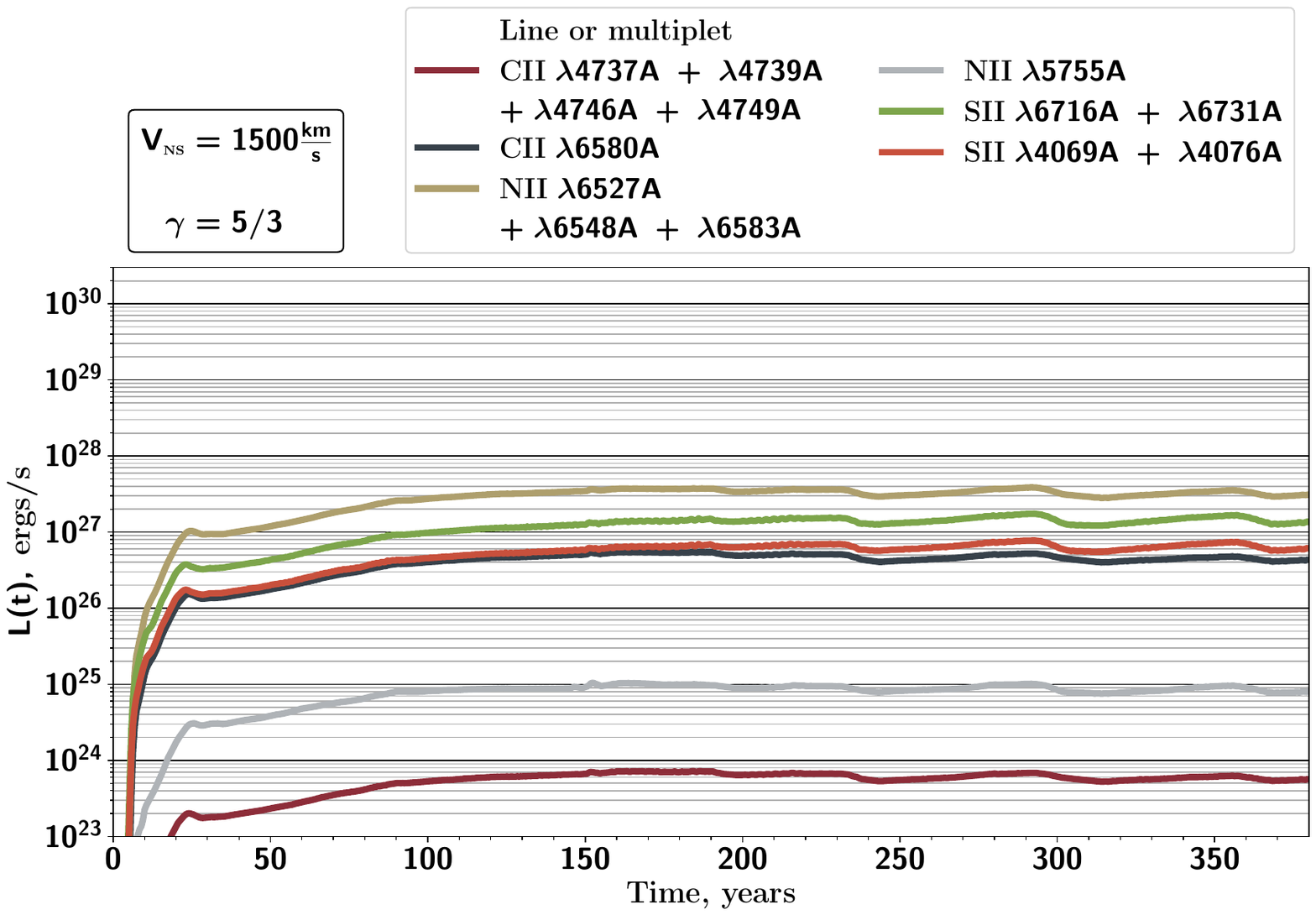}
        \includegraphics[width=.49\textwidth, angle=-0, trim=0 0 0 0]{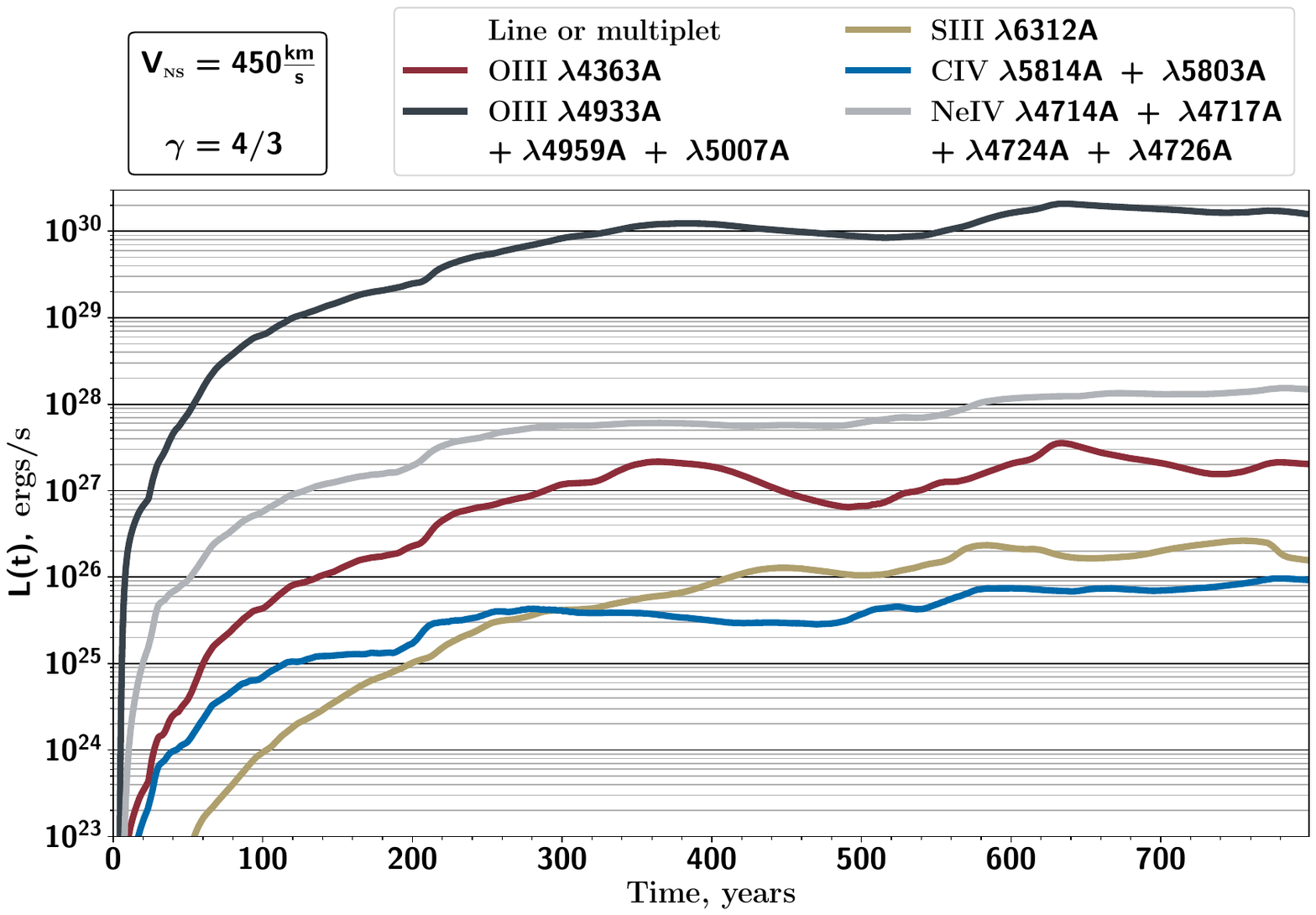}
        \includegraphics[width=.49\textwidth, angle=-0, trim=0 0 0 0]{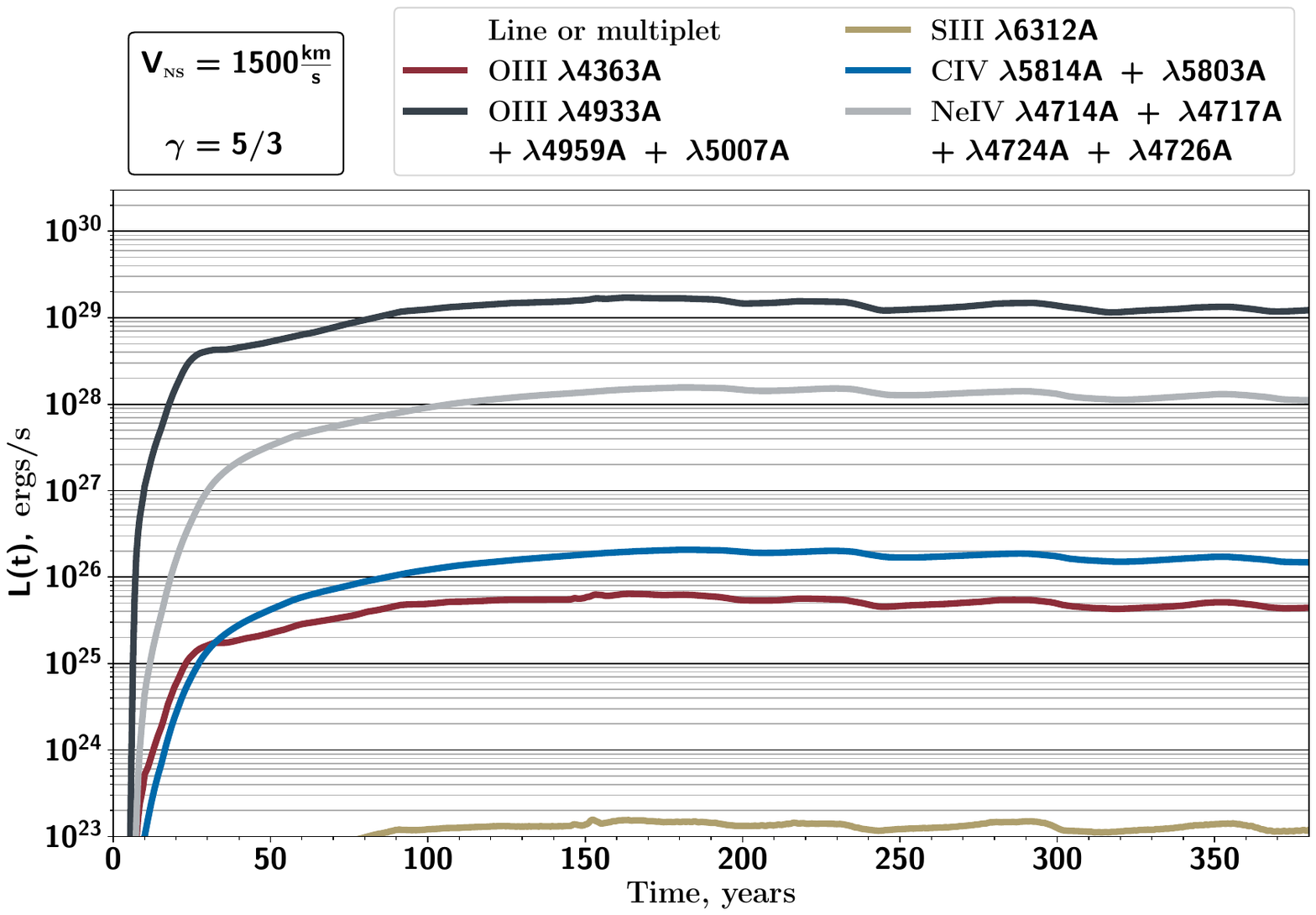}
        \caption{Light curves of nebula in v03g43 and v1g53 (the brightest and the dimmest rings, respectively). Lines are divided into groups by ionisation stages and shown on different plots.}        
        \label{fig:linesLightCurves}
\end{figure*}

Here, we plot light curves of model nebulae following equation:
\begin{equation}
        L (t) = \int_{z_{min}}^{z_{max}} dz \int_{0}^{R_{max}} 4 \pi \eta \left( (R,z), t \right) 2 \pi R dR.
\end{equation}
The example light curves are provided in Figure~\ref{fig:linesLightCurves}.

At the start of the integration, there is no partially ionised gas in the model, so the luminosity in different lines is zero. During the simulation, the quasistationary ionisation regime on the bow shock settles, which leads to rapid rise of luminosity until it reaches the plateau. This happens relatively fast in the case of \Ha\ and slower for lines of elements in higher ionisation stages.

When the quasistationary solution is achieved, there is some variability on light curves. The reason is the difference between the shape of model nebulae and Mach cone. The rise of luminosity is due to rings emerging, enlarging, and thus carrying more material. Then luminosity rapidly falls back to the plateau, when the ring exits computational domain.

\end{document}